# Alfvén Waves in Solar Flares

*Alexander Russell, University of Dundee, Scotland, UK*
*Recently moved to University of St Andrews, Scotland, UK*

12 September 2023

**Abstract:** Solar flares are dramatic events in which magnetic reconnection in the corona leads to heating of plasma to tens of MK and acceleration of particles to high energies. They also centrally involve transport between the corona (where the magnetic reconnection occurs) and the lower solar atmosphere (where most energy is radiated from). There is substantial evidence for the presence of Alfvénic waves/turbulence in solar flares, for example, in the ubiquitous nonthermal broadening of flare spectral lines. The physical role that Alfvénic waves have in the flare has attracted considerable attention, especially since 2007–2010. This article reviews what spectroscopic observations reveal about the properties and importance of Alfvénic waves, turbulence and transport in solar flares; mechanisms for wave excitation by magnetic reconnection at high Lundquist numbers and braking of the sunward reconnection jet; and models of wave energy transport to the lower atmosphere and the resulting heating and dynamics. The article finishes with discussion of the outlook for new progress.

Table of Contents

# 1. INTRODUCTION

Solar flares are spectacular centerpieces of astrophysics that combine natural drama with the important impacts of space weather. For the scientifically curious, they possess a physical richness that interleaves diverse astrophysical plasma processes, span environments from the solar interior to the solar wind and leave their mark across the full range of the electromagnetic spectrum. They are a prime example of a plasma phenomenon for which scientific advance requires not only greater understanding of components that are enigmatic and challenging of themselves, but also integration of components into the complete system.

This review examines the role of Alfvénic waves in solar flares. We hope to provide an overview of major points in this rapidly maturing field, including: observational constraints on the locations, timings and importance of Alfvénic waves in solar flares; generation of waves/turbulence by magnetic reconnection at high Lundquist numbers and braking of the sunward reconnection jet; and downward transport to the lower solar atmosphere and the ensuing dynamics there. The article finishes with discussion of the outlook for new progress and a summary.

This article is part of an interdisciplinary collection on Alfvén waves in the heliosphere and it is important to highlight at the outset that the investigation of Alfvén waves in solar flares differs from other plasma environments. In environments blessed with in-situ observations, such as planetary magnetospheres and the solar wind, extensive data sets directly confirm the presence of Alfvén waves, establish key attributes such as wavelength and Poynting flux, document variations under different conditions, and connect these properties with phenomena such as particle energization and auroral emission. In contrast, detection of waves in solar environments relies on remote observation, which has been sufficiently challenging that imaging of waves in the solar corona was achieved only at the end of the 20$^{th}$ Century. Continued advances have since established that magnetohydrodynamic (MHD) waves are ubiquitous in the Sun's corona and lower atmosphere, and sophisticated seismological techniques are used to infer local plasma properties from wave observations. There is also compelling evidence that upgoing MHD waves make important contributions to heating magnetically open parts of the corona and accelerating the fast solar wind. These topics are reviewed in other articles in this collection.

For flares, remote observations show beyond doubt that solar flares excite MHD waves in their surroundings. As early as 1960, Hα observations revealed the occurrence of Moreton waves in the chromosphere, which spread out from the flare site like a blast wave (Moreton and Ramsey 1960). Later, coronal imaging found the coronal counterpart in the form of EUV waves (see recent discussion about interpretation of these observations by Long et al. (2017)). Flare-induced shock waves are implicated in the generation of Type II radio bursts (Uchida 1960) and further evidence that flares launch disturbances into their surroundings is provided by the excitation of transverse kink oscillations of nearby coronal loops (Nakariakov et al. 1999; Aschwanden et al. 1999). Finally, wave oscillations are a leading candidate to explain quasi-periodic pulsations in flare light curves and radio emission, although slower compressible wave modes, such as slow modes and sausage modes, are typically considered responsible rather than Alfvén waves (see reviews by Nakariakov and Melnikov (2009) and McLaughlin et al. (2018)).

Solar flares also produce rapid, significant and persistent changes of the photospheric magnetic field (Wang et al. 1994; Cameron and Sammis 1999; Kosovichev and Zharkova 2001). These photospheric field changes occur after the start of the flare, and they are believed to be a response to the restructuring of the coronal magnetic field. Sudol and Harvey (2005) documented many cases where components of the photospheric magnetic field changed by more than 0.01 T in less than ten minutes (also see additional studies by Petrie and Sudol (2010) and Petrie (2013), and the review by Wang and Liu (2015)). Alfvénic wave fronts are a natural agent for communicating the magnetic field changes, motivating consideration of large-amplitude Alfvénic waves in flares (Fletcher and Hudson 2008). In this context, resonant coupling between the downgoing Alfvén wave front and acoustic waves may excite seismic waves in the solar interior known as sunquakes (Hudson, Fisher, and Welsch 2008; Wang and Liu 2010; Fisher et al. 2012; Russell et al. 2016). Furthermore, Johnstone, Petrie, and Sudol (2012) found an observational connection between photospheric magnetic field changes and enhanced chromospheric UV emissions, consistent with Alfvénic fronts heating the chromosphere on their way to the underlying photosphere.

Unfortunately, directly observing Alfvén waves at the heart of the flare is challenging for several fundamental reasons. One factor is time-resolution: an upper limit on wave periods is the duration of the impulsive phase of a solar flare (defined in Section 2), which is typically of the order of several minutes. More challengingly, high cadence radio and X-ray observations vary on timescales as short as tens of milliseconds (Benz 1986; Kiplinger et al. 1983). Imaging high-frequency waves is clearly more challenging than resolving the 3-to-5-minute periods typical of waves in the background Sun, as faster cadences starve instruments of photons. Another limitation is spatial resolution: as a benchmark on current capabilities, the *High-resolution Coronal Imager* (Kobayashi et al. 2014; Rachmeler et al. 2019), which has an upcoming flare mission planned for 2024, has a pixel resolution of approximately 0.13″, or about 100 km on the Sun. Coronal waves with scales shorter than this cannot currently be spatially resolved. Finally, solar flares create challenging conditions for coronal observation such as bright backgrounds and large dynamic range. These issues can be mitigated by observing strategies such as restricting attention to low-energy flares, or limb flares that are partially occulted by the solar disk; but they nonetheless provide significant practical constraints.

For a long time, Alfvén waves and Poynting flux lay outside the main set of concepts used in flare models. Instead, debate centered on whether energy was delivered to the lower atmosphere by thermal conduction (the so-called "thermal" flare model) or electron beams (the "non-thermal" flare model). There were nonetheless thinkers willing to point out how strange it would be if solar flares were a rare exception to the universal importance of Poynting flux in astrophysical plasmas, and if the burstiness of the flare energy release did not excite waves. Further encouragement to consider Alfvén waves and Poynting flux could be taken from nonthermal broadening of flare spectral lines, magnetic field changes, the persistence of the electron number and supply problems in the collisional thick-target model (Fletcher and Hudson 2008; Brown et al. 2009), and analogies to other systems such as Earth's magnetosphere, in which the importance of wave transport was becoming increasingly clear.

The intellectual conditions changed dramatically during 2007–2010, which we attribute to a confluence of events. First, 2007 was a breakthrough year for observations of transverse waves in the solar corona and the tipping point for the narrative that MHD waves are "ubiquitous" in the corona, which quickly became consensus in solar physics (Tomczyk et al. 2007; De Pontieu et al. 2007; Lin et al. 2007; Okamoto et al. 2007). Second, during 2008–2010, simulations of magnetic reconnection crossed the Biskamp criterion (Lundquist number $S_L > 10^4$) beyond which an initial Sweet-Parker reconnection layer is disrupted by plasmoid instability (Lapenta 2008; Bhattacharjee et al. 2009; Huang and Bhattacharjee 2010; Daughton et al. 2011). It rapidly became clear that magnetic reconnection at high Lundquist numbers is fast (normalized reconnection rates are 0.01–0.1 and independent of $S_L$) and highly dynamic. This development gave fresh impetus to the point long made by flare observations that the coronal energy release is highly variable, hence likely to launch waves. It also ushered in a new era in which some aspects of wave generation by magnetic reconnection can be probed using simulations. Third, increasingly sophisticated spaceborne spectrometers provided novel opportunities to identify where and when unresolved plasma motions occur in solar flares: *Hinode*/EIS (Culhane et al. 2007) launched 22 September 2006 and *IRIS* (De Pontieu et al. 2014) launched 27 June 2013; together these missions have substantially advanced knowledge of waves/turbulence in solar flares.

We also highlight two major papers in this period. "Impulsive Phase Flare Energy Transport by Large-Scale Alfvén Waves and the Electron Acceleration Problem" by Fletcher and Hudson (2008), was especially influential and it ignited interest in wave energy transport among flare researchers. The paper "Chromospheric Evaporation by Alfvén Waves" by Haerendel (2009), while less frequently cited initially, also played an important role in launching this topic, introducing many of the ideas that will feature in this review.

The last 15 years have been an exciting period for the study Alfvénic waves in solar flares, during which the topic has matured rapidly. Today, many of the central issues and tools have been clearly identified and new advances are being driven by advances of simulation and observing capabilities. As we hope the article will convey, this is an exciting time to study the role of Alfvénic waves in solar flares and we believe that many opportunities lie ahead.

## 2. SOLAR FLARES BACKGROUND

As a general definition, solar flares are events in which magnetic energy stored in the Sun's corona is converted to enhanced radiative emissions. The energies involved can be very large, in extreme events exceeding $10^{26}$ J ($10^{33}$ erg). The bulk of the radiated energy is normally emitted from the lower solar atmosphere in the white light and infrared parts of the spectrum; however, the pre-flare energy storage and release occur in the corona, so flare physics is closely associated with coronal processes and energy transport. Furthermore, flares can produce emissions across the electromagnetic spectrum, including radio, microwave, infrared, white light, extreme ultraviolet (EUV), X-rays and gamma rays. An observational overview can be found in the article by Fletcher et al. (2011).

Since 1966 the International Astronomical Union has classified solar flares according to the peak X-ray flux measured by satellite observatories in Earth orbit (in the 1–8 Å range). The scale employed for this purpose employs scientific notation, but uses letters X, M, C, B and A to denote the magnitude (respectively $10^{-4}$, $10^{-5}$, $10^{-6}$ $10^{-7}$ and $10^{-8}$ W m$^{-2}$). For example, an X8.2 flare is one for which the peak X-ray flux at Earth was $8.2 \times 10^{-4}$ W m$^{-2}$.

Flares are diverse phenomena: they exhibit power law distributions across broad ranges of energies; occur in active regions of different magnetic morphologies and complexities; and happen throughout the life cycle of individual active regions. Flare light curves exhibit several types of evolution, e.g., impulsive, gradual or long duration; and flares can be, but are not always, associated with coronal mass ejections (CMEs). To streamline discussion, we focus on the canonical flare evolution of an impulsive phase followed by a gradual phase.

Observationally, the impulsive phase corresponds to a rapid rise in the 1–8 Å X-ray flux, during which impulsive phenomena are detected such as hard X-rays, microwave emission and explosive upflows of plasma from the lower atmosphere. The impulsive phase is associated with the primary energy release, whereby magnetic reconnection liberates energy from the coronal magnetic field. A large portion of the released energy is ultimately transported to the lower solar atmosphere, forming flare ribbons and driving upflows known as "chromospheric evaporation" that fill coronal loops with relatively dense plasma that radiates at high temperatures. The impulsive phase typically concludes within ten minutes, and it is followed by a longer-lasting gradual phase, during which the X-ray emission decreases from its peak as the flare loops cool by radiation and thermal conduction. Timescales suggest there is continued heating during the gradual phase, but it is less intense than during the impulsive phase.

The observational features of solar flares are intimately related to the topology of the coronal magnetic field. While different flares exhibit various morphologies, this article will consider the CSHKP "standard" model of an eruptive two-ribbon flare (Carmichael 1964; Sturrock 1966; Hirayama 1974; Kopp and Pneuman 1976). This phenomenology is especially important for space weather, since it is common among the largest X-class flares. It is also relevant to many of the flare observations, simulations and theoretical works discussed in this article.

The CSHKP model proposes that magnetic reconnection occurs in an extended current sheet in the corona, building on earlier ideas by Sweet (1958). This configuration is sketched in Figure 1. Energy released by magnetic reconnection is transported along the magnetic field, forming flare ribbons in the chromosphere that are magnetically connected to the coronal current sheet and/or reconnection outflow jet. This resembles auroral energy deposition near the magnetic footpoints of the reconnection site in a planetary magnetotail.

We also draw attention to the "above-the-loop" region that lies between the top of the flare loops and the bottom of the current sheet. The magnetic field line sketch in Figure 1 suggests the existence of a region of low magnetic field strength (field strength is inversely proportional to field line spacing), which coincides with where the sunward reconnection outflow jet is forced to brake. This region was not emphasized historically; however, given that flow braking typically redirects large-scale bulk motion into small-scale turbulence (*c.f.* the pool below a waterfall, or the interface region around supernova remnants) and the pocket of weak magnetic field forms a magnetic trap for particles, we believe that the above-the-loop region should be of great physical interest.

The above-the-loop region has also become interesting for observational reasons, starting with the detection of a hard X-ray source at this location (Masuda et al. 1994; Petrosian, Donaghy, and McTiernan 2002; Sui and Holman 2003; Krucker and Lin 2008; Liu et al. 2008; Ishikawa et al. 2011). Microwave observations analyzed by Fleishman et al. (2022) found a magnetic field strength of 200–400 G in this region, compared to kilogauss field strengths elsewhere in the corona. Various investigations have found that almost all the electrons in this region are accelerated during flares, creating a nonthermal population with a number density of $10^{10}$–$10^{11}$ cm$^{-3}$ (Krucker et al. 2010; Krucker and Battaglia 2014; Fleishman et al. 2022; Chen et al. 2020). This density is comparable to the electron beam density that would be required to power radiative emissions from flare ribbons in powerful events (Krucker et al. 2011; Ishikawa et al. 2011), although the particle flux to the lower atmosphere depends on what fraction of particles are in the magnetic trap's loss cone.

The 2D version of the CSHKP standard flare model shown in Figure 1 is a sufficient guide to follow the rest of this article. We emphasize, though, that the true dynamics occur in 3D, which has two major implications. First, even when the time-averaged dynamics are 2D, the 3D nature of the physical processes involved is crucial, as will be seen when discussing the generation of Alfvén waves in Section 6. Second, the global magnetic configuration is 3D, and readers interested in the 3D aspects of the global topology are directed towards Moore and Labonte (1980), Moore et al. (1997), Aulanier, Janvier, and Schmieder (2012), Amari, Canou, and Aly (2014), Janvier et al. (2014), Janvier, Aulanier, and Démoulin (2015) and references therein.

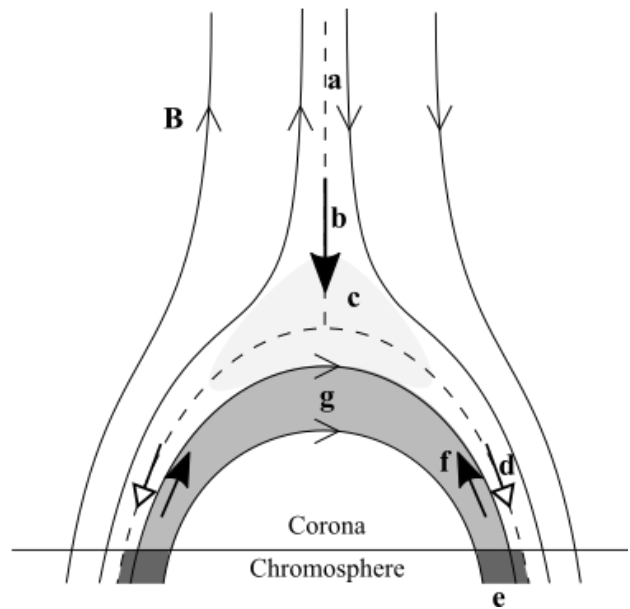

Figure 1 – CSHKP model of a two-ribbon flare. Magnetic field lines sketch the magnetic geometry, with dashed lines representing magnetic separatrices. Features include: **a** Reconnecting current sheet; **b** Sunward reconnection outflow (speed approximately $v_A$); **c** Above-the-loop region (light grey) associated with braking of the sunward reconnection outflow, particle trapping and nonthermal electrons; **d** Field-aligned energy transport (unfilled arrow heads) to the chromosphere by electron beams, thermal conduction, Alfvénic waves or a combination thereof; **e** Flare ribbons (bright emission from the heated chromosphere); **f** Chromospheric evaporation (dense upflows from the chromosphere to the corona); **g** Flare loops in X-rays and EUV, which are bright having been filled by evaporation.

## 3. TERMINOLOGY: ALFVÉN AND ALFVÉNIC

It is worthwhile to explain what we mean by Alfvén waves vs. Alfvénic waves. As is well known, a homogeneous equilibrium supports three MHD wave modes: the Alfvén, fast and slow modes. The Alfvén wave, discovered by Alfvén (1942), is characterized by field-aligned transport at the Alfvén speed,

$$v_A = \frac{B}{\sqrt{\mu_0 \rho}}.$$

It is a transverse wave, with magnetic field perturbations perpendicular to the background magnetic field. The restoring force is magnetic tension, and it is incompressible (unlike the magnetoacoustic modes). Finally, propagating Alfvén waves satisfy the Walén relation,

$$\frac{\delta \boldsymbol{B}}{B_0} = \mp \frac{\delta \boldsymbol{v}_\perp}{v_A},$$

where the negative (positive) sign is for Alfvén waves propagating parallel (antiparallel) to the background magnetic field. In space physics, the Walén relation is often expressed as the ratio of the electric field perturbation to the magnetic field perturbation equaling the Alfvén speed. It is also a statement that the Alfvén wave's kinetic and magnetic energies are equipartitioned.

Alfvén waves also exist as nonlinear solutions for homogeneous equilibria. Finite-amplitude circularly polarized Alfvén waves, which have uniform magnetic pressure, retain $v_{||} = 0$ and are strictly incompressible. On the other hand, finite-amplitude linearly polarized Alfvén waves perturb the magnetic pressure, which leads to "ponderomotive" flows of plasma, such that $v_{||} \neq 0$. This nonlinear Alfvén wave is no longer strictly incompressible, and it produces density perturbations that travel with the wave. This outcome illustrates that one must beware of inadvertently introducing fragility into classification schemes for MHD waves by insisting too strongly on idealized properties such as $\nabla \cdot \boldsymbol{v} = 0$, which may not transfer to more realistic situations such as non-vanishing wave amplitudes.

In solar physics, a prominent debate around terminology (e.g., Van Doorsselaere, Nakariakov, and Verwichte 2008; Goossens et al. 2009) has stemmed from the need to consider inhomogeneous equilibria, most pressingly dense structures aligned with the magnetic field known as coronal loops. Magnetic cylinder models of the type examined by Edwin and Roberts (1983) support linearized wave modes that are distinct to those of uniform medium, of which, kink waves and azimuthal Alfvén waves are most pertinent to the present discussion. The azimuthal Alfvén waves are polarized such that velocity perturbations lie in flux surfaces of constant density, and they behave similarly to the Alfvén mode for a uniform medium. They include the "torsional Alfvén wave", which is strictly incompressible in these models and has an azimuthal wave number of $m = 0$, but this class can also be extended to include higher azimuthal wave numbers, for which the waves are weakly compressible.

The kink wave is a mode of the inhomogeneous structure, and it relies on small departures from $\nabla \cdot \boldsymbol{v} = 0$, which make it qualitatively distinct from a so-called "pure" Alfvén wave. Nonetheless, kink and Alfvén waves share many properties. Most importantly, in the slender tube limit (i.e. when parallel wavelengths are much greater than the tube radius), kink waves are transverse waves that transport energy along the magnetic field at the kink speed,

$$c_k = \sqrt{\frac{\rho_0 v_A + \rho_e v_{Ae}}{\rho_0 + \rho_e}},$$

where $\rho_0$ and $v_A$ are the mass density and Alfvén speed inside the cylinder and $\rho_e$ and $v_{Ae}$ are their counterparts outside the cylinder. Inspecting the formula above, one sees that the kink speed is a mean of the Alfvén speeds in the cylinder and its environment, which bound its value. The wave is therefore said to be "Alfvénic" (Edwin and Roberts 1983). Also in common with Alfvén waves, kink waves are only weakly compressible in the corona, and the dominant restoring force is magnetic tension (Goossens et al. 2009). We further add that (considering for simplicity the case of a homogeneous equilibrium magnetic field) the linearized induction equation implies the following generalization of the Walén relation for propagating waves:

$$\frac{\delta \boldsymbol{B}_\perp}{B_0} = -\frac{\delta \boldsymbol{v}_\perp}{c},$$

where $c$ is the wave speed, set to $\pm c_k$ for kink waves compared to $\pm v_A$ for Alfvén waves. Hence, in both cases, the fluctuation ratio yields an Alfvénic speed.

It is helpful at this point to define a class of "Alfvénic waves" with similar properties, of which both "pure" Alfvén waves and kink waves are members. Properties this class of MHD waves are:

- Field-aligned transport at Alfvénic speeds.
- Perturbations predominantly transverse to $\boldsymbol{B}_0$.
- Low compressibility.
- Magnetic tension the dominant restoring force.
- Propagating waves have $\delta \boldsymbol{B}_\perp / B_0 = -\delta \boldsymbol{v}_\perp / c$, where $c$ is the (Alfvénic) phase speed.

The semantics and history of "Alfvén" vs. "Alfvénic" are discussed in depth in Section 2 of the article by Magyar in this collection. In agreement with what we have written here, Magyar concludes: "The term Alfvénic reflects the fact that transverse waves in a generally inhomogeneous plasma of parallel wavelengths longer than the typical perpendicular scale of inhomogeneity have dominantly Alfvén-like properties. In this sense, it is the term that can most broadly describe transverse waves in an inhomogeneous plasma, without making implicit assumptions about the geometry of the inhomogeneity."

It is clearly desirable to make a precise distinction as to whether a transverse wave is a "pure" Alfvén wave or a kink wave, when there is sufficient information on the equilibrium density and the perturbations to do so. This is usually achievable for coronal loops that persist throughout the period of study. In other situations, remote observations are ambiguous as to the existence and extent of transverse density structuring. Furthermore, at the heart of a solar flare, the notion of a gently perturbed equilibrium is distinctly questionable, but MHD simulations of turbulence, reconnection and turbulent magnetic relaxation show that Alfvénic fluctuations are ubiquitous even under such conditions. We therefore believe it is often appropriate to identify a wave as Alfvénic without attempting finer distinctions.

Finally, one might ask why not simply name this broader class Alfvén waves? Doing so could, for example, better align with terminology such as "poloidal Alfvén waves" used in magnetospheric physics. Here, we can only answer that it is not the custom in solar physics to do so.

In this article, sections addressing observations will typically refer to Alfvénic waves, which are taken to be a class of MHD waves that have the properties itemized above. Occasionally, we refer to this general class as Alfvén waves, as in the article's title – we do this sparingly and to provide consistency with other fields that do not use the "Alfvénic" terminology, acknowledging the interdisciplinary nature of the collection to which this article belongs. When using mathematical models to explore issues such as energy transport and wave dissipation, we usually consider "pure" Alfvén waves for concreteness, which is indicated in the text by naming them specifically as Alfvén waves.

## 4. NONTHERMAL LINE WIDTHS

The first question to ask about Alfvénic waves in solar flares is "Do flare generated Alfvénic waves leave a fingerprint in solar observations?" In answer, we spotlight nonthermal broadening of spectral lines in solar flares. If excess spectral line widths in flares can be attributed to transverse waves, one might reasonably expect these observations to advance diagnostics of Alfvén waves in solar flares, similarly to how hard X-ray diagnostics have advanced the study of flare nonthermal electrons.

### 4.1 Spectral Broadening by Unresolved Plasma Motions

A spectral line emitted by stationary plasma has a line width determined by Doppler broadening due to random thermal motions of the particles. Departures provide information about unresolved plasma motion – a principle that has long been exploited by astrophysicists and solar physics.

In the simplest models, nonthermal broadening is evaluated using the formula

$$(\Delta\lambda)_{nth} = \sqrt{(\Delta\lambda)^2_{obs} - (\Delta\lambda)^2_{inst} - (\Delta\lambda)^2_{th}},$$

where $(\Delta\lambda)_{obs}$ characterizes the observed line profile, $(\Delta\lambda)_{inst}$ is the instrumental broadening, and $(\Delta\lambda)_{th}$ is the line width in the absence of unresolved motions.

Estimating the thermal width $(\Delta\lambda)_{th}$ requires the ion temperature. One of the more precise methods for obtaining a temperature is to infer the electron temperature from intensity ratios of temperature-sensitive spectral lines, then assume thermal equilibrium between species such that $T_i \approx T_e$. Allowing for the possibility that $T_i \neq T_e$, some investigators have used the alternative simple heuristic of assuming the ion temperature is close to the peak formation temperature of the line they are using, although this assumption carries significant uncertainty since ionization states are populated for a broad range of temperatures around the peak formation temperature. A more sophisticated approach uses Differential Emission Measure (DEM) analysis, which combines information from multiple ionization states.

Once obtained, nonthermal widths can be converted to velocity units using the Doppler formula

$$v_{nth} = c \frac{(\Delta\lambda)_{nth}}{\lambda_0},$$

where $\lambda_0$ is the rest wavelength of the line center. The resulting $v_{nth}$ provides a representative magnitude for unresolved line-of-sight velocities.

Alternatively, the nonthermal velocity can be expressed as

$$v_{nth} = \sqrt{\frac{2k_B}{m_i}(T_D - T_e)},$$

where the Doppler temperature of the emitting ion is defined by

$$\sqrt{\frac{2k_B T_D}{m_i}} = \frac{c}{\lambda_0}\sqrt{(\Delta\lambda)^2_{obs} - (\Delta\lambda)^2_{inst}}.$$

Finally, we point out that it is not always so simple. The "sum-of-squares" formulas used above are rigorous for gaussian line profiles, which are common in flare spectra, but not universal. Instrumental quirks (such as convolution with the instrumental spread function) or plasma conditions (such as non-Maxwellian ion populations) can produce non-gaussian line profiles. In such cases, $v_{nth}$ typically becomes a curve fitting parameter. It is also common that line widths are reported using the full width at half-maximum (FWHM) instead of fitting the gaussian width, which introduces a factor $2\sqrt{\ln(2)} \approx 1.665$ in some papers.

### 4.2 Whole-active-region X-ray Spectra

X-ray spectroscopy of the Sun developed rapidly after World War II, as solar scientists developed payloads for rocket, satellite and space station programs, including OSO 6 (Doschek et al. 1971), Intercosmos-4 (Grineva et al. 1973) and Skylab (Doschek et al. 1975). These spectrometers did not spatially resolve features within flaring active regions, but they nonetheless provided considerable insights into the properties of flare plasma. The occurrence of nonthermal line broadening during impulsive flares was well established by the late 1970s, and it was a major scientific interest for the second-generation X-ray spectrometers that flew aboard *P78-1* (launched 24 Feb 1979), *Solar Maximum Mission* (*SMM*; launched 14 Feb 1980) and *Hinotori* (launched 21 Feb 1981). Doschek (1990) provides a general discussion of X-ray flare spectra results from this period. A decade later, *Yohkoh* (launched 30 Aug 1991) continued these investigations using an X-ray spectrometer roughly an order of magnitude more sensitive than previous instruments.

The main properties of nonthermal broadening were already known by the time of the systematic series of studies by Doschek, Kreplin, and Feldman (1979), Doschek et al. (1980) and Feldman et al. (1980), which analyzed data from the SOLFLEX Bragg crystal spectrometers on the *P78-1* satellite. These observations covered a sample of M and X class flares, with a focus on the Fe XXV and Ca XIX lines. The main conclusions about line widths were:

1. Line profiles produced in hot flare plasma (temperatures ≥ 10 MK) display nonthermal broadening of hundreds of kilometers per second, up to ~160 km s$^{-1}$ in the SOLFLEX data.

2. The broadening is greatest during the impulsive phase, which is when Fe XXV lines are usually first detected. Nonthermal widths decrease steadily from the initial peak.

3. Nonthermal velocities in the gradual phase are considerably lower than during the impulsive phase but remain elevated, with ~60 km s$^{-1}$ being common.

4. The magnitude of nonthermal motions is a weak function of temperature. Hotter lines have larger nonthermal velocities.

These conclusions were confirmed by *SMM* (Gabriel et al. 1981; Culhane et al. 1981; Antonucci et al. 1982; Antonucci, Gabriel, and Dennis 1984) and *Hinotori* (Tanaka et al. *1982)*. In many observations, the peak $v_{nth}$ was recorded when counts first became sufficient to fit the line; thus, the true maximum almost certainly occurred when the Fe XXV line was too faint to be reliably fitted. Since greater sensitivity allows measurements earlier in the flare, the best estimates of the peak $v_{nth}$ in whole-active-region spectra came from *Yohkoh* BCS. We therefore adopt the peak $v_{nth}$ for looptop sources of 250 km s$^{-1}$ from Khan et al. (1995), consistent with Tanaka et al. (1982) and Mariska, Sakao, and Bentley (1996), rather than the SOLFLEX value quoted in Point 1. Figure 2 shows plots by Alexander et al. (1998) that illustrate Points 1 to 4 in two flares observed by *Yohkoh*.

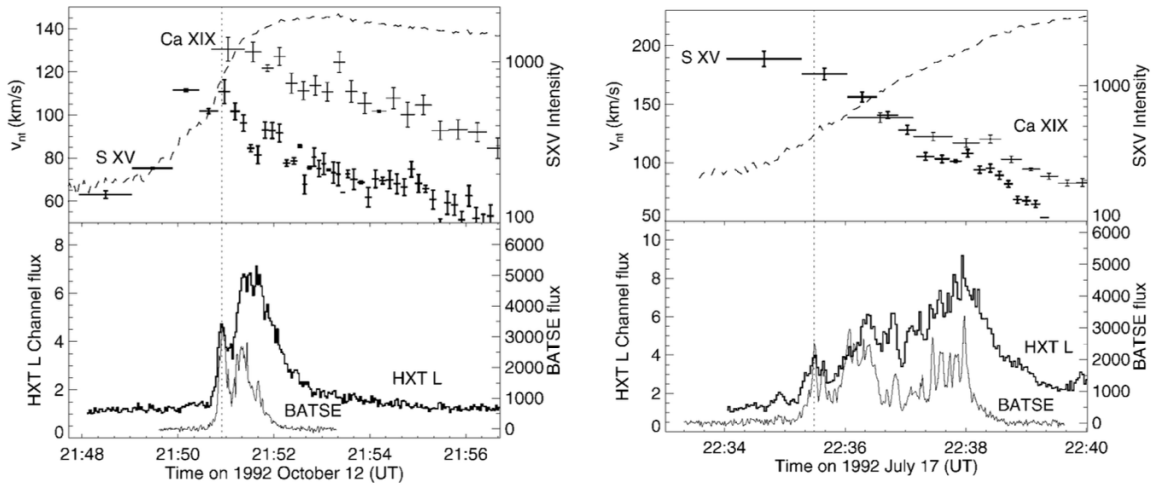

Figure 2 – Evolution of flare nonthermal velocities (top panels, markers, left axis) with comparison to line intensity (top panels, dashed lines, right axis) and hard X-ray emission (bottom panels), for two solar flares observed by Yohkoh. Nonthermal widths of hundreds of km s$^{-1}$ are seen minutes before the HXR peak. Reproduced from Alexander et al. (1998). © AAS. Reproduced with permission.

Since nonthermal broadening is observed early in the flare, it is interesting to explore the timing of its peak relative to other impulsive phase phenomena. Timing studies by Antonucci, Gabriel, and Dennis (1984), Alexander et al. (1998) (see Figure 2), Harra, Matthews, and Culhane (2001) and Ranns et al. (2001) found evidence that:

5. Nonthermal widths can increase a few minutes or more *before* the onset of impulsive phenomena such as the hard X-ray burst and appearance of high-speed upflows, although they are not always seen to do so.

The likely cause of nonthermal broadening is also informed by comparisons across different events. Using *SMM* data, Antonucci et al. (1982) and Antonucci (1984) found blueshifts of 300 to 400 km s$^{-1}$ occurred for disk flares but were absent for limb flares, consistent with bulk upward radial flows at a significant fraction of the sound speed. In contrast, it was found that:

6. The nonthermal velocity does not show significant center-to-limb variation.

These center-to-limb dependencies were confirmed by statistical surveys with *Yohkoh,* covering more than 100 events (Mariska, Doschek, and Bentley 1993; Mariska 1994). Points 5 and 6 together provide evidence that a significant part of the nonthermal broadening is not due to superposition of emission from upflows at multiple speeds.

A final major result from whole-active-region studies is the finding by Fludra et al. (1989) that:

7. The nonthermal and upflow velocities are correlated in most flares.

Points 5 and 7 suggest that the nonthermal broadening is connected to the primary energy release.

### 4.3 Spatially Resolved EUV Observations – General Remarks

We now turn to observations that resolve specific locations within active regions, moving to lines in the extreme-ultraviolet (EUV). There is considerable heritage to EUV spectroscopy, which includes instruments aboard *SMM* and the *Solar and Heliospheric Observatory* (*SOHO*). In the last decade, the instruments that have arguably contributed most to new knowledge are *Hinode*/EIS (Culhane et al. 2007) and *IRIS* (De Pontieu et al. 2014).

Spectral coverage for EIS includes lines of Fe XXIV and Fe XXIII, while IRIS coverage includes Fe XXI. Applying the trend that peak nonthermal velocities are lower in cooler lines, one should expect nonthermal velocities observed by EIS and IRIS to be lower than for Fe XXV lines in X-ray spectra. In other words, one should bear in mind that EIS and IRIS do not have access to the hottest flare plasma, which is where the strongest motions seem to occur.

Single-slit spectrographs can create raster "images" by scanning the slit position. It is important to be aware that pixel values in different stripes of a raster correspond to different times. The rastering process can also be slow, leading to long effective cadences. These considerations mean that rasters are better suited to studying the spatial distribution of $v_{nth}$ during the phase in which nonthermal velocities are decaying, than they would be for searching for the peak value of $v_{nth}$, which is short-lived. In contrast, the stare mode of operation keeps the slit pointed at one location. Sometimes, the slit is at an interesting location, in the best cases from before the start of the impulsive phase. *IRIS* has the advantage that as a slit-jaw spectrometer it captures contextual slit-jaw images that show exactly where the spectra correspond to.

Looking to the future, the *Multi-slit Solar Explorer* (*MUSE*; De Pontieu et al. 2020) is designed to simultaneously obtain EUV spectra along 37 slits, allowing observation of a 170″ × 170″ field of view with a cadence around 1 s and raster time of 12 s. Furthermore, its spectral coverage is well suited to flare studies (Cheung et al. 2022). We anticipate that this future mission will have a large scientific impact on the study of nonthermal broadening and Alfvén wave/turbulence in solar flares, a point we return to in Section 8.2.

Sections 4.4 to 0 summarize current knowledge from spatially resolved EUV spectra. They are organized according to spatial location, starting with observations of the flare plasma sheet, followed by above-the-loop and looptop sources, and finishing with the flare footpoints. Interpretation of the nonthermal broadening is discussed in Section 4.7.

### 4.4  Flare/CME Plasma Sheets

The reconnecting current sheet is a central feature of the standard flare model (Figure 1) and associated plasma sheets are sometimes imaged behind CMEs. Clear views are relatively rare, requiring viewing the plasma sheet end-on, but several cases have been documented. As elaborated below, spectroscopic measurements reveal these sheets to be a site of significant nonthermal broadening, especially at early times in the eruption.

*SOHO*'s Ultraviolet Coronal Spectrometer (UVCS) investigated several post-CME plasma sheets, typically for heliocentric distances between 1.5 and 2.0 $R_\odot$ and using Fe XVIII emission. The "Halloween Event" (4 November 2003) study by Ciaravella and Raymond (2008) has the distinction of measuring the line widths at relatively early times. Their results detected $v_{nth} \approx 380$ km s$^{-1}$ for their earliest data point, which was followed by rapid decay over ten minutes. A few hours later, $v_{nth}$ around 50 to 100 km s$^{-1}$ is typical in post-CME plasma sheets (Ciaravella et al. 2002; Ciaravella and Raymond 2008; Bemporad 2008; Schettino, Poletto, and Romoli 2010; Susino, Bemporad, and Krucker 2013).

More recently, *Hinode*/EIS has also investigated nonthermal broadening within post-CME plasma sheets. For the Textbook Flare on 10 September 2017, EIS collected raster spectroscopic images that extend from the solar limb to approximately 1.24 $R_\odot$ heliocentric altitude. Nonthermal velocities of up to 200 km s$^{-1}$ for Fe XXIV were detected, displaying the typical behavior of being strong early in the flare and decreasing thereafter (Warren et al. 2018; Li et al. 2018; Cheng et al. 2018; French et al. 2020).

The spatial distribution of $v_{nth}$ found by Warren et al. (2018) and Li et al. (2018) is interesting too. First, $v_{nth}$ increases with altitude in the plasma sheet. Second, the profile of $v_{nth}$ perpendicular to the plasma sheet appears single-peaked at higher altitudes; but at lower altitudes it appears double-peaked, with ridges of $v_{nth}$ coinciding with where one would expect magnetic separatrices as the end of the current sheet opens in a cusp. These observations are consistent with reconnection models in which the strongest turbulence is generated in the vicinity of a dominant reconnection site above the maximum altitude of the EIS observations and turbulence/waves exiting from the end of the current sheet propagate near the magnetic separatrices. Alternatively, the double-ridge structure may indicate generation of small-scale motions by instabilities, for instance, Kelvin-Helmholtz instability fed by the velocity shear at the edge of the outflowing reconnection jet (Loureiro, Schekochihin, and Uzdensky 2013).

## 4.5 Above-the-loop Region and Loops

We now move down to the above-the-loop regions associated with coronal hard X-ray and microwave emission. This region is interesting from the perspectives of wave/turbulence generation by braking of the reconnection outflow jet and particle acceleration, and it displays some of the largest nonthermal velocities outside the plasma sheet.

Figure 3 shows a limb flare analyzed by Shen et al. (2023). This region displays the canonical $v_{nth}$ evolution of steady decline from an initial peak greater than ~ 100 km s$^{-1}$. Similar $v_{nth}$ values for the above-the-loop region have been reported for different flares by Tian et al. (2014) and Polito et al. (2018).

Covering a larger field of view, Figure 4 shows a nonthermal velocity map from the 15 May 2013 X1.2 flare (Doschek, McKenzie, and Warren 2014; Kontar et al. 2017; Stores, Jeffrey, and Kontar 2021). Blue contours show *SDO*/AIA 94 Å intensity, indicating flare loops; white contours show *SDO*/AIA 304 Å intensity, indicating flare ribbons; and the orange pixels show the Fe XXIV nonthermal velocity obtained from an EIS raster. This map supports several important conclusions:

1. The greatest nonthermal velocity is at the apex of the above-the-loop region.

2. The value of $v_{nth}$ decreases by around 50% as one descends towards the footpoints.

3. Ridges of $v_{nth}$ are located further out than the AIA 94 Å loops.

Point 3 is consistent with the broadening being associated with energy transport from the reconnection site in the corona. It might also indicate $v_{nth}$ is greatest in the hottest loops, which occur further out than the brightest loops (Tsuneta 1996). All three points are further evidence that the nonthermal broadening is related to the primary energy release, not upflows from the chromosphere.

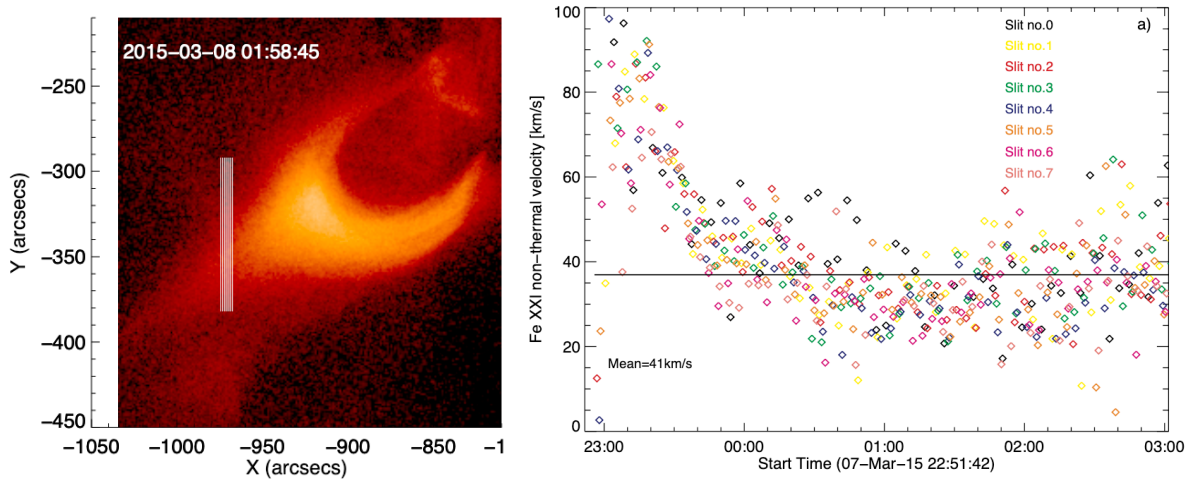

Figure 3 – Evolution of nonthermal velocities in the above-the-loop region of a long-duration M9.2 solar flare. The left panel is an Hinode/XRT context image that shows the locations sampled by IRIS during the narrow raster. The right panel shows the time development of nonthermal velocities in this region. Reproduced from the arXiv version of Shen et al. (2023) under a CC BY 4.0 license.

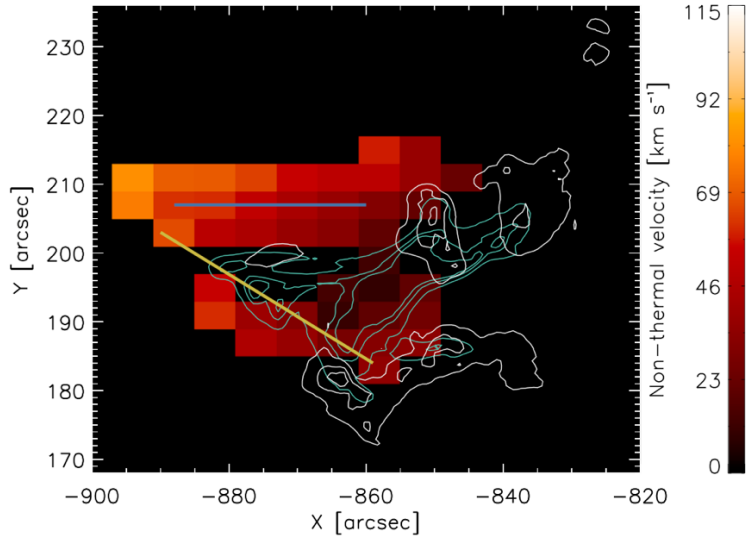

Figure 4 – Map of nonthermal velocity in the corona during an X1.2 flare. Orange pixels show nonthermal velocity measured from the EIS Fe XIV line. Cyan contours show AIA 94 Å intensity, outlining flare loops. White contours show AIA 304 Å intensity highlighting flare ribbons. Reproduced from Stores, Jeffrey, and Kontar (2021) under a CC BY 4.0 license.

We also highlight a few earlier works of this type. In one of the first EIS studies, Hara et al. (2008) mapped line widths for a C2 class long duration flare near the limb. Their results demonstrated that Ca XVII line widths are largest at the top of the above-the-loop region, and widths of the cooler Fe X and Fe XII lines are greatest at the top of the bright flare loops. The nonthermal broadening was consistent with motions at 50 to 100 km s$^{-1}$. Hara et al. (2011) subsequently studied Fe XXIII and Fe XXIV during the impulsive phase of a B9.5 long duration flare near disk center, finding nonthermal broadening of ~100 km s$^{-1}$ at the first timepoint, steadily decreasing thereafter. Harra et al. (2013) produced maps of $v_{nth}$ for four near-limb flares of M and X classifications. Three of the flares investigated by Harra et al. (2013) were eruptive and displayed pre-flare enhancements of $v_{nth}$ in the above-the-loop region. The largest eruptive flare (X2.1) was viewed close to the limb and values of $v_{nth}$ in the above-the-loop region displayed an initial peak between 100 and 150 km s$^{-1}$, decaying to less than 50 km s$^{-1}$. Doschek, McKenzie, and Warren (2014) analyzed four near-limb flares, including the 15 May 2013 X1.2 flare shown in Figure 4. They also analyzed the long-duration X1.7 flare on 27 January 2012, which was viewed at an orientation presenting an arcade of flare loops in the plane of sky. For this latter event, Doschek, McKenzie, and Warren (2014) measured plane-of-sky velocities by applying local correlation tracking (LCT) to AIA 131 Å images, which are dominated by Fe XXI emission. The LCT analysis revealed velocity variations by 100 to 200 km s$^{-1}$ over distances of just a few megameters (also see, McKenzie 2013). This result reinforces the view that nonthermal broadening in the above-the-loop region is indeed due to plasma motions. Finally, we mention that flare nonthermal velocities around 100 km s$^{-1}$ have also been reported using *SOHO*/SUMER (Landi et al. 2003) and *SOHO*/CDS (Brosius 2012).

## 4.6 Footpoints

Whole-active-region X-ray spectra for flares near disk center often have two main components: one stationary and the other blue-shifted (Antonucci et al. 1982; Antonucci, Gabriel, and Dennis 1984; Fludra et al. 1989; Doschek et al. 1989; Antonucci, Dodero, and Martin 1990). It is reasonable to identify the stationary component with the looptop sources discussed in the previous section. The blueshifted source is believed to be located at the magnetic footpoints and associated with chromospheric evaporation. The nonthermal widths of the blueshifted and stationary components can be significantly different (Doschek et al. 1989; Fludra et al. 1989).

Following earlier work using *SOHO*/CDS (Brosius 2003), EIS made it possible to study footpoint sources across a broad temperature range with high spatial and spectral resolution. For example, Milligan and Dennis (2009) and Milligan (2011) studied EIS spectra for the footpoints of a C1.1 flare, reporting nonthermal broadening coincident with hard X-ray footpoint sources. Their results are summarized by Figure 5. Cooler lines (≤ 4 MK) exhibit modest blueshifts and redshifts (50 km s$^{-1}$ or less) and the nonthermal widths of 55–80 km s$^{-1}$, with no clear relation between their nonthermal widths and temperatures. The hottest lines, Fe XXIII and Fe XXIV (10–30 MK), had profiles consistent with a dominant stationary component plus a strongly blueshifted component. The blueshifted components exhibited upward velocities of around 250 km s$^{-1}$ and fitted nonthermal widths were reported to be around 100–120 km s$^{-1}$. Young et al. (2013) found similar results for a flare kernel during the rise phase of a M1.1 class flare, confirming the occurrence of nonthermal broadenings of 100 to 120 km s$^{-1}$ in Fe XXIII and Fe XXIV, with significantly less broadening of cooler lines.

The temperature dependence of the Doppler velocity is an expected result, since 1D hydrodynamic simulations show that rapid deposition of energy in the chromosphere causes an explosive upflow, accompanied by a momentum-conserving cool downflow at lower altitudes (Fisher, Canfield, and McClymont 1985). The enhancement of nonthermal velocity for temperatures above 10 MK indicates that $v_{nth}$ is significantly larger for the explosively upflowing plasma than it is deeper in the chromosphere, which would be consistent with waves dissipating in the heated layer.

An important advance since Milligan (2011) has been the resolving of footpoint sources. Brosius (2013) reported a completely blueshifted Fe XXIII (~14 MK) footpoint source for one C1 class flare with EIS. Soon after, *IRIS* made the significant breakthrough of regularly spatially resolving Fe XXI (~10 MK) at the footpoints, obtaining totally blueshifted profiles (Tian et al. 2014; Tian et al. 2015; Young, Tian, and Jaeggli 2015; Graham and Cauzzi 2015; Li et al. 2015; Polito et al. 2015; Polito et al. 2016; Lee et al. 2017). When Fe XXI is first detected, the line is typically blueshifted by 100 to 400 km s$^{-1}$, sometimes taking it to the edge of the instrument's spectral read-out window, and it has a large nonthermal width, often around 100 to 120 km s$^{-1}$. Over the next several minutes, the blueshift steadily decreases to zero, and the nonthermal broadening decreases to around 50 km s$^{-1}$ or less.

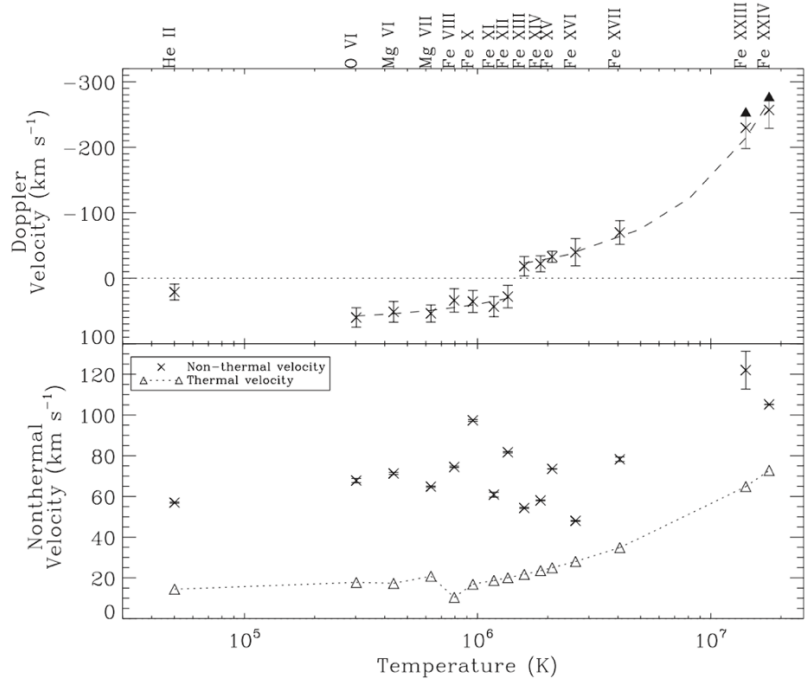

Figure 5 – Line-of-sight Doppler velocity (top panel) and nonthermal velocity (bottom panel) in footpoints of a C1.1 flare observed by Hinode/EIS. Hot lines (10–30 MK) display nonthermal widths up to 120 km s$^{-1}$, compared to 55–80 km s$^{-1}$ for cooler lines (≤ 4 MK). Reproduced from Milligan et al. (2011). © AAS. Reproduced with permission.

If the footpoint broadening is interpreted as due to Alfvénic waves, then the hot flare lines provide the velocity amplitude above where the chromosphere is heated. A limitation is that sufficient Fe XXI intensity to fit the line is produced tens of seconds after the presumed energy deposition (see Figure 3 of Graham and Cauzzi (2015)). As for whole-active-region X-ray spectra, high temperature EUV lines provide a lower bound for the maximum $v_{nth}$, but the true maximum of $v_{nth}$ almost certainly occurs before there is sufficient intensity to fit the lines.

To gain insight into times before and during the onset of flaring in individual pixels, one can turn to lines that exist at chromosphere and transition region temperatures. Jeffrey et al. (2018) investigated this question for a B class flare using *IRIS* Si IV (~80,000 K). The evolution is plotted in Figure 6 (top panel) at 1.7 s cadence. The Si IV non-thermal velocity starts at a pre-flare value of about 9 km s$^{-1}$, increases over ten seconds to a maximum of 30 km s$^{-1}$, and returns to preflare levels within a minute. The increase in $v_{nth}$ leads the increase in the *IRIS* Si IV and *RHESSI* 6 to12 keV X-ray intensities by about 20 s. Also shown (bottom panel) is the position of the line center and the corresponding Doppler velocity of the line shift (negative values are blueshifts). Jeffrey et al. (2018) suggested their observations are consistent with unresolved plasma motions in the flare ribbons heating the plasma with a dissipation time of ~10 s.

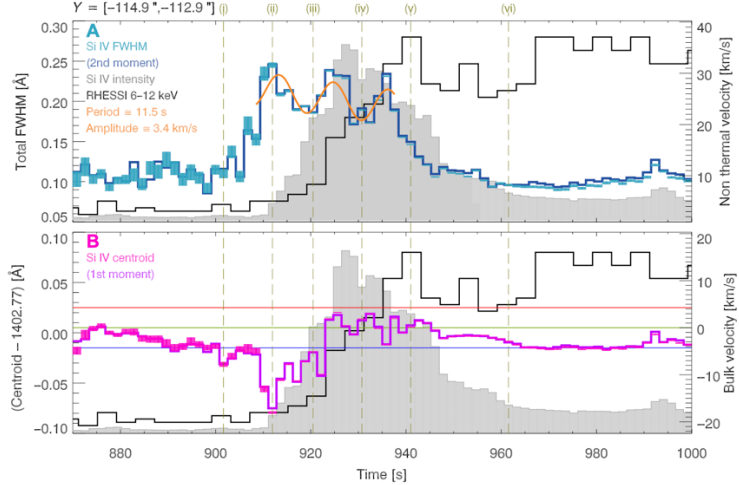

Figure 6 – Evolution of Si IV in the footpoint of a B-class flare. The top panel (A) shows the total line width (left axis) and the equivalent $v_{nth}$. The line width increases from its preflare value of 10 km s$^{-1}$ to a peak of 30 km s$^{-1}$. This increase in nonthermal broadening leads the increases of the Si IV intensity and 6–12 keV X-rays by 10–20 s. The bottom panel (B) shows the evolution of the line centroid. Reproduced from Jeffrey et al. (2018) under a CC BY 4.0 license.

The timings might tentatively connect the nonthermal broadening with footpoint heating before the impulsive phase (also see Fletcher et al. (2013) and Simões, Graham, and Fletcher (2015)). Analysis of spatial scales along the ribbons for the same flare by French et al. (2021) have added to the timeline. Their work showed that an instability, hypothesized to be tearing instability of the coronal current sheet, first develops at a single "key scale", then spreads to smaller and larger scales, ending in fully developed turbulence with a power law slope of 2.3. The rise in nonthermal widths reported by Jeffrey et al. (2018) coincides with the stage in which the turbulence is developed across multiple scales. Additional examples of *IRIS* Si IV nonthermal broadening preceding flaring have been reported by Chitta and Lazarian (2020).

### 4.7  Interpretation

The consensus explanation for nonthermal broadening is unresolved mass motions. Various other line broadening mechanisms exist, such as pressure broadening, opacity broadening and the Stark effect, e.g., discussion by Milligan (2011). The difficulty for many of the alternative hypotheses is the ubiquity of nonthermal broadening during flares, which occurs in the flare plasma sheet, above-the-loop region, loops and ribbons, and for a multitude of optically thin lines.

One simple alternative to unresolved motions would be that the line width reveals the true ion temperature, which differs from the electron temperature and the expected formation temperature of the line. This hypothesis has been previously rejected, with Antonucci (1984) writing that, "The observed profiles would imply a difference in ion and electron temperatures sometimes exceeding one order of magnitude. At the density inferred for the thermal plasma of flares, it would be difficult to explain the persistence of large differences in electron and ion temperatures over the observed flare rise times." The inferred ~100 MK ion temperatures would also be many times the peak formation temperature of the observed lines. This hypothesis therefore appears less plausible than unresolved mass motions.

Several studies have looked at the role of non-Maxwellian particle distributions, including Jeffrey, Fletcher, and Labrosse (2016), Dudík et al. (2017) and Polito et al. (2018). Broad EUV spectral lines in flares often have gaussian profiles, which are consistent with a Maxwellian ion distribution; however, a kappa profile provides a better fit in some cases. Dudík et al. (2017) concluded that kappa distributions do not remove the presence of non-thermal broadening, but on the contrary imply larger unresolved motions.

Support evidence that line broadening is due to unresolved motions can be found in direct imaging of plasma motions. Local correlation tracking (LCT) has been applied to *SDO*/AIA 131 Å images, finding variations in local velocities by hundreds of km s$^{-1}$ over distances of a few megameters (Doschek, McKenzie, and Warren 2014; McKenzie 2013). Meanwhile, MHD simulations have demonstrated generation of small-scale plasma motions of 100 km s$^{-1}$ or greater in the braking or "interface" region, which propagate to the chromospheric footpoints as Alfvénic waves. When line profiles are forward modelled for these simulations, they produce broad gaussian profiles that match observations (Shen et al. 2023; Ruan, Yan, and Keppens 2023). These simulations are discussed further in Section 6.3.

An historical debate about the form of the unresolved motions has been between a superposition of field-aligned flows, or motions that are either isotropic or perpendicular to the magnetic field. Arguments against attributing broadening to superposition of field aligned flows have been made since at least the 1980s: a simple model involving vertical upflows is inconsistent with the center-to-limb behavior; upflows struggle to explain line widths in loop-top sources; and such models cannot account for events in which nonthermal broadening precedes occurrence of upflows. As spectral and spatial resolutions have improved, the symmetric broad profiles of resolved observations have added further weight against the idea that broadening is due to a mixture of upflows (Polito, Testa, and De Pontieu 2019). Timings further make it likely that broadening is related to the energy release itself, rather than evaporative upflows.

Linear polarization (Stokes *P*) also provides information relevant to this debate. For a resolved magnetic field, *P* depends on the angle between the magnetic field and the radial direction, and the inferred magnetic field direction typically agrees with expectations from intensity images. Examining the 10 September 2017 Textbook Flare using the Coronal Multi-channel Polarimeter (CoMP) instrument, French et al. (2019) also found that *P* drops to low values in the plasma sheet and in the above-the-loop region, which is where large nonthermal broadening is detected in this flare (Warren et al. 2018; Li et al. 2018). The reduction in linear polarization signal was attributed to unresolved magnetic fields with mixed orientations. Putting together the bigger picture, the unresolved velocities (nonthermal broadening) are accompanied by unresolved magnetic field orientations (low values of *P*). This association would not occur for field-aligned flows, but it is consistent with large-amplitude Alfvénic waves.

This article takes the view that the inferred motions most likely have an Alfvénic nature. Maps of $v_{nth}$ in the corona (Figure 4) are consistent with field-aligned transport, as is the location of broadening at the flare footpoints. The transport is rapid and therefore better associated with the Alfvén speed than the sound speed. The unresolved motions are accompanied by unresolved variations in the direction of the magnetic field, indicating that they are not hydrodynamic in nature. If one ascribes unresolved motions to turbulence, fluctuations in strongly magnetized plasma typically have an Alfvénic character. At the same time, further work is required to conclusively establish whether unresolved motions are primarily perpendicular to the background magnetic field, e.g., viewing angle studies at modern spatial resolution.

## 5. WAVE ENERGY DENSITY AND POYNTING FLUX

Combining the values of $v_{nth}$ in previous sections with a few assumptions and physical properties, one can estimate the energy density and Poynting flux of the inferred Alfvén waves. For these calculations, we assume that magnetic fluctuations are perpendicular to the background magnetic field, and there is equipartition of wave energy between kinetic and magnetic energies.

The averaged wave energy density can be expressed as

$$\langle E_{wave} \rangle = \frac{m_i n \langle v^2 \rangle}{2} + \frac{\langle \delta B^2 \rangle}{2\mu_0} = m_i n \langle v^2 \rangle.$$

Coronal densities for flaring solar active regions are typically around $10^{17}$ m$^{-3}$, so approximating the coronal $m_i$ as the proton mass and considering $v_{nth} \approx 250$ km s$^{-1}$ for the above-the-loop region, the energy density of the waves is estimated as $\langle E_{wave} \rangle \approx 10$ J m$^{-3}$.

For comparison, the thermal energy density of the ions (or electrons) is

$$E_{th} = \frac{n k_B T}{\gamma - 1}.$$

Setting the adiabatic index as $\gamma = 5/3$ and considering the number density used above, a temperature of 10 MK gives $E_{th} \approx 21$ J m$^{-3}$. In these estimates, the wave energy density is a significant fraction (about half) of the thermal energy, and high enough to play a significant role in flare energetics, for example by accelerating particles in the above-the-loop region. This is especially true if the waves are acting as an intermediary via which energy in converted to other forms, as discussed by Kontar et al. (2017).

The comparison of energy densities can also be performed by expressing the ratio as

$$\frac{\langle E_{wave} \rangle}{E_{th}} = \frac{(\gamma - 1) m_i \langle v^2 \rangle}{k_B T} = \gamma(\gamma - 1) \frac{\langle v^2 \rangle}{c_s^2},$$

which makes clear that the ratio is independent of the density. Here, $c_s = \sqrt{\gamma p / \rho} = \sqrt{\gamma k_B T / m_i}$ is the ion sound speed. From this formulation, we see that the wave energy density is a significant fraction of the thermal energy density because the $v_{nth}$ is a significant fraction of the ion sound speed.

We are also interested in the ability of waves to transport energy to the lower atmosphere. Poynting's theorem (with the energy density of the electric field neglected for non-relativistic MHD) states that

$$\frac{\partial}{\partial t}\left(\frac{B^2}{2\mu_0}\right) + \nabla \cdot \boldsymbol{S} = -\boldsymbol{j} \cdot \boldsymbol{E},$$

such that magnetic energy is transported by the Poynting vector

$$\boldsymbol{S} = \frac{1}{\mu_0} \boldsymbol{E} \times \boldsymbol{B}.$$

We will estimate the Poynting flux for an Alfvén wave, assuming the ideal Ohm's law

$$\boldsymbol{E} = -\boldsymbol{v} \times \boldsymbol{B}.$$

Assuming that magnetic and velocity perturbations are perpendicular to the background magnetic field, and focusing on transport parallel to the background magnetic field, we are interested in

$$\boldsymbol{S}_\parallel = \frac{1}{\mu_0}(\boldsymbol{B}_0 \times \delta\boldsymbol{v}) \times \delta\boldsymbol{B} = -\frac{1}{\mu_0}(\delta\boldsymbol{B} \cdot \delta\boldsymbol{v})\boldsymbol{B}_0.$$

Finally, eliminating $\delta\boldsymbol{B}$ using the Walén relation for propagating Alfvén waves,

$$\frac{\delta\boldsymbol{B}}{B_0} = \mp \frac{\delta\boldsymbol{v}}{v_A},$$

the averaged parallel Poynting flux is

$$\langle S_\parallel \rangle = \frac{B_0^2}{\mu_0 v_A}\langle v^2 \rangle = v_A m_i n \langle v^2 \rangle = v_A \langle E_{wave} \rangle = B_0 \sqrt{\frac{m_i n}{\mu_0}}\langle v^2 \rangle.$$

This result can also be obtained heuristically using the principle that the wave energy density is transported at the Alfvén speed.

Magnetograms reveal active region magnetic field strengths of up to several kG in the photosphere. In the active regions, microwave observations and magnetic field extrapolations also imply kG field strengths in the corona (Brosius and White 2006). For the above-the-loop region, field strengths are lower, e.g., around 300–400 G (0.03–0.04 T) for the 10 September 2017 Textbook Flare (Fleishman et al. 2022) and the 15 May 2012 flare (Kontar et al. 2017).

We now estimate $\langle S_\parallel \rangle$ for the apex and footpoints, setting the number density of $10^{17}$ m$^{-3}$ for both cases. For the above-the-loop region, a magnetic field strength of 0.04 T and velocity amplitude of 250 km s$^{-1}$ yield a Poynting flux of $2.9 \times 10^7$ W m$^{-2}$ ($2.9 \times 10^{10}$ erg s$^{-1}$ cm$^{-2}$). In this case, the Alfvén speed is 2800 km s$^{-1}$, and $\langle E_{wave} \rangle \approx 10$ J m$^{-3}$ as calculated earlier in this section. Near the footpoint, a magnetic field strength of 0.1 T and velocity amplitude of 120 km s$^{-1}$ yield a Poynting flux of $1.7 \times 10^7$ W m$^{-2}$. For this case, the Alfvén speeds is 6900 km s$^{-1}$ and $\langle E_{wave} \rangle \approx 2.4$ J m$^{-3}$.

The Poynting fluxes calculated above exceed $10^7$ W m$^{-2}$ and they are just slightly below the solar luminosity per unit area of $7 \times 10^7$ W m$^{-2}$. Both estimates are consistent with the normal range of energy deposition rates considered in flare modelling, which is $10^6$ to $10^8$ W m$^{-2}$ (Fisher, Canfield, and McClymont 1985). The estimated Poynting fluxes are also comparable to electron energy deposition rates for specific events (Antonucci, Dodero, and Martin 1990; Lee et al. 2017) and broadband white light and UV emissions (Fletcher and Hudson 2008). Various 1D radiative-hydrodynamic simulations, including the wave-driven model of Reep and Russell (2016), have shown that energy fluxes of $10^7$ W m$^{-2}$ are capable of driving flare-like dynamics, including strong footpoint heating and explosive chromospheric evaporation. We therefore conclude that if nonthermal broadening is attributed to Alfvénic waves, the waves carry enough energy to contribute meaningfully to flare dynamics.

There are some events for which nonthermal electrons have been estimated to carry substantially larger energy fluxes into the flare footpoints than the values discussed above. Specifically, Krucker et al. (2011) derived an energy deposition rate by electron beams in the region of $5 \times 10^9$ W m$^{-2}$ for the brightest footpoints during an X6.5 flare. While that estimate is sensitive to the low-energy cutoff assumed for the electron beam, and this flare appears to be an outlier compared to more common observational estimates of order $10^7$ W m$^{-2}$, it poses the question of whether Alfvén waves could plausibly deliver a similar Poynting flux in extreme circumstances.

The available parameters are the number density, magnetic field strength and wave amplitude. As a reference point for number density, Milligan (2011) determined a footpoint density of $10^{17.5}$ m$^{-3}$ for a C1.1 class flare, which is slightly higher than the density value we used above, for a relatively low energy flare. Meanwhile, 1D radiative hydrodynamics simulations show that large energy deposition rates produce dense chromospheric evaporations, with densities around $10^{19}$ m$^{-3}$ not uncommon. We might therefore be justified in considering a footpoint density of $10^{19}$ m$^{-3}$ for an X6.5 flare with energy deposition rate of $5 \times 10^9$ W m$^{-2}$. Increasing the assumed density by two orders of magnitude (e.g., from $10^{17}$ m$^{-3}$ to $10^{17}$ m$^{-3}$) increases the Poynting flux by one order of magnitude. Similarly, given this is an extreme event, the footpoint magnetic field strength might be raised to 2 kG and the peak value of $v_{nth}$ to 250 km s$^{-1}$, noting Point 7 in Section 4.2 that stronger upflows correlate with larger nonthermal widths. With these modifications, the Poynting flux would be $1.4 \times 10^9$ W m$^{-2}$, which is comparable to the energy deposition rate derived by Krucker et al. (2011). This exercise, while speculative, suggests that Alfvénic waves can deliver sufficient energy fluxes to contribute meaningfully to the dynamics of even the most extreme flares.

The main conclusion of this section is that observationally derived parameters for $v_{nth}$, density and magnetic field strength imply that the wave energy density at the apex is comparable to the ion thermal energy, and Poynting fluxes are large enough to meaningfully affect the lower atmosphere (typically several times $10^7$ W m$^{-2}$). The effects of this energy transport are explored in Section 7.

## 6. GENERATION OF ALFVÉNIC WAVES/TURBULENCE

This article now moves to theory and simulations, starting with the generation of Alfvénic waves in solar flares. A major shift in flare theory during the last two decades has been the pivot from laminar steady-state models to dynamic ones in which the magnetic field and flow may be nonlaminar. In this respect, theory has been catching up with the empirical fact that solar flares are highly dynamic with variability on sub-second timescales (Kiplinger et al. 1983; Benz 1986). Here we focus on two promising wave sources: dynamic magnetic reconnection and braking of the sunward reconnection outflow. Our main goals are to understand qualitatively what processes are responsible for generating the waves/turbulence observed in solar flares, and to constrain the likely spectra of frequencies and wavenumbers.

As context, we note that many features of the Sweet-Parker model of reconnection (Parker 1957a; Sweet 1958) apply well to the CSHKP solar flare model shown in Figure 1, in particular the magnetic geometry, and Alfvénic outflows from the ends of the current sheet. Despite those successes, the resistive, steady-state and laminar Sweet-Parker model faces several well-known problems. The best-known one concerns the reconnection rate, which Sweet-Parker theory gives as $v_{in}/v_A \sim S_L^{-1/2}$, where

$$S_L = \frac{v_A L}{\eta}$$

is the Lundquist number based on the half-length of the current sheet, $L$, and $\eta$ is the magnetic diffusivity. A typical solar Lundquist number is $S_L \sim 10^{12}$, for which classic Sweet-Parker reconnection predicts a reconnection rate of order $10^{-6}$. In contrast, observations of solar flares indicate reconnection rates are of order 0.01–0.1. A second problem is that solar flares exhibit a "switch-on" effect, in which the impulsive phase starts suddenly following preflare evolution. Steady state models like classic Sweet-Parker or Petschek reconnection (Petschek 1964) are, by definition, unable to address this triggering.

The dynamic reconnection models that we discuss in connection with wave generation (in Sections 6.1 and 6.2) are fast (their reconnection rates are 0.01–0.1 consistent with flare observations) and they switch on (via tearing instability). As of recently, there is no longer a rate or triggering problem, although major work is ongoing to understand 3D reconnection at high Lundquist numbers, in MHD and kinetic frameworks.

## 6.1 Plasmoid Mediated Reconnection

It has been known for many decades that Sweet-Parker layers are unstable to tearing instability (Furth, Killeen, and Rosenbluth 1963) above a critical Lundquist number $S_c$. When $S_L < S_c$, the Sweet-Parker flow stabilizes the tearing instability (Bulanov, Syrovatskiĭ, and Sakai 1978), whereas for $S_L > S_c$ tearing instability develops. The canonical instability threshold is $S_c \sim 10^4$ (Biskamp 1986), although some researchers have suggested it might be as low as $10^3$.

While several of the key ideas in the theory of plasmoid-mediated reconnection were developed before 2008 (Biskamp 1986; Shibata and Tanuma 2001; Loureiro, Schekochihin, and Cowley 2007), a sea change occurred when computing power made it possible to simulate Sweet-Parker layers with $S_L > 10^4$ in 2.5D. This threshold was crossed when an MHD simulation by Lapenta (2008) demonstrated the switch-on of a fast reconnection regime at the critical Lundquist number. The fast reconnection was subsequently confirmed by other investigators and theoretical understanding quickly developed (Bhattacharjee et al. 2009; Cassak, Shay, and Drake 2009; Huang and Bhattacharjee 2010; Uzdensky, Loureiro, and Schekochihin 2010). The papers by Huang and Bhattacharjee (2013) and Loureiro and Uzdensky (2016) offer detailed overviews.

Of interest to wave generation, the nonlinear evolution for $S_L > S_c$ produces a chain of plasmoids separated by reconnecting current sheets. This configuration is highly dynamic, with plasmoids constantly being created, growing, moving, merging and being expelled from the reconnection layer. Current layers within the plasmoid chain are on average marginally stable, such that if they are Sweet-Parker layers then $S_l \sim S_c$ yields a reconnection rate of $S_c^{-1/2} \sim 0.01$, where the lower case subscript $l$ is used to distinguish the half-length of the marginally stable Sweet-Parker layers from the half-length $L$ of the global reconnection layer. This is consistent with MHD simulation results. If the widths of the smallest current are thinner than the ion inertial length or gyroradius, they exhibit kinetic reconnection, yielding a faster global reconnection rate of order 0.1 (Birn et al. 2001; Daughton et al. 2009; Shepherd and Cassak 2010).

From a wave-generation perspective, we take a special interest in the fact that the global outflows from plasmoid-mediated reconnection exhibit large time variability. The outflows are not continuous; they are instead "lumpy" because of the expulsion of discrete plasmoids. Distributions of plasmoid sizes have been investigated analytically and numerically, with Fermo, Drake, and Swisdak (2010), Uzdensky, Loureiro, and Schekochihin (2010), Loureiro et al. (2012) and Huang and Bhattacharjee (2012) having developed the kinetic theory of plasmoids. The theoretical distribution function of plasmoids with respect to flux is a power law $f(\psi) \sim \psi^{-1}$, falling off more steeply for the largest plasmoids of which there is typically only one at any given time (Huang and Bhattacharjee 2012, 2013).

2.5D simulations have been produced that explore plasmoid-mediated reconnection directly in a solar flare context. The simulation by Karpen, Antiochos, and DeVore (2012) forms a CSHKP current sheet self-consistently via the magnetic breakout mechanism, which evolves gradually until it becomes tearing unstable, switching on fast plasmoid-mediated reconnection. During the fast reconnection phase, plasmoids are frequently ejected from the current sheet in the sunward outflow and collide with the underlying closed magnetic loops, exciting disturbances that propagate along the magnetic field towards the lower atmosphere (e.g., animated version of their Figure 13). Burstiness of outflows is also seen in flare observations, e.g., Cheng et al. (2018).

The frequent but irregular expulsion of flux ropes with a distribution of sizes leads us to expect the excitation of waves with a corresponding distribution of wavelengths and frequencies. For a solar flare, the distribution of plasmoid widths, and hence wavelengths, is expected to span from the effective thickness of the dynamic reconnection layer based on the fluctuation EMF (see, Shibata and Tanuma 2001; Beg, Russell, and Hornig 2022) down to the ion scales. For the Textbook Flare, this range of wavelengths corresponds to 10 Mm or greater at the outer scale (Li et al. 2018), whereas ion scales in the corona are of the order of meters. For timescales, we expect the largest flux ropes to be ejected at intervals comparable to the Alfvén travel time along the lower altitude half of the flare current sheet, setting an outer period $\tau_{max} \sim L/v_A$. We tentatively connect this outer period to the timescale of elementary flare bursts, which are typically several seconds to a few tens of seconds in duration (de Jager and de Jonge 1978). Superimposed would be variability determined by the interval between smaller plasmoids, which occur more frequently according to the $f(\psi) \sim \psi^{-1}$ distribution. The shortest timescale that emerges is the length of a marginally stable reconnecting current sheet divided by the Alfvén speed, which yields $\tau_{min} \sim S_c \eta / v_A^2$. For coronal values, $\tau_{min}$ can be shorter than milliseconds. Spikes and other subsecond variability observed in flares (Kiplinger et al. 1983; Benz 1986) might therefore be related to expulsion of flux ropes of a variety of sizes from the flare current sheet.

## 6.2  Reconnection Layer Turbulence

The real Sun is of course 3D, and the additional dimensional freedom qualitatively changes the nature of magnetic reconnection compared to 2.5D. Foremost, magnetic field lines are not constrained to lie on flux surfaces but can instead be stochastic. Field line wandering is the crucial ingredient in the turbulent reconnection paradigm advanced by Lazarian and Vishniac (1999), and it modifies particle acceleration because particles are no longer trapped in magnetic islands (Dahlin, Drake, and Swisdak 2015, 2017). Furthermore, stochasticity arises naturally in reconnection because flux ropes formed by tearing instability undergo secondary 3D instabilities such as kink instability, which break any initial 2.5D symmetry. Knowledge of wave generation during reconnection should therefore come from 3D models.

Early numerical tests of reconnection in the presence of turbulence relied on artificially driven turbulence (Matthaeus and Lamkin 1986; Loureiro et al. 2009; Kowal et al. 2009). However, there is now a substantial collection of 3D simulations that show that when $S_L > S_c$, turbulence is self-generated and self-sustaining within the reconnection layer. For example, MHD simulations by Oishi et al. (2015), Huang and Bhattacharjee (2016), Striani et al. (2016), Beresnyak (2017), Kowal et al. (2017, 2020), Yang et al. (2020), Beg, Russell, and Hornig (2022), and Daldorff, Leake, and Klimchuk (2022); and kinetic simulations by Daughton et al. (2011), Liu et al. (2013), Daughton et al. (2014), Stanier et al. (2019), Zhang et al. (2021), and Li et al. (2019).

The zoo of 3D self-generated turbulent reconnection (SGTR) simulations is still being systematically organized. Simulations that explicitly include global reconnection outflows and have a guide field comparable to the reconnecting magnetic field component, such as Huang and Bhattacharjee (2016) and Beg, Russell, and Hornig (2022), typically form dynamic frayed flux ropes like those shown in Figure 7. These structures appear to make reconnection fast by a plasmoid-mediated mechanism. In such cases, generation of Alfvénic waves by the processes discussed in Section 6.1 is likely to carry over to 3D. Additionally, a new source of Alfvénic waves in 3D is that reconnection is expected to form magnetic twists locally, which propagate as Alfvénic waves. Since the magnetic field is stochastic over longer distances (notice the mixing of field lines towards the z boundaries in Figure 7), the stochastic phase mixing (Similon and Sudan 1989) described in Section 7.3 will transfer wave energy to shorter perpendicular scales within the reconnection layer.

Simulations with periodic boundaries along the current layer typically behave differently, with the turbulent layer thickening over time and a reconnection process that bears greater resemblance to Lazarian and Vishniac (1999). We suspect those latter model setups are less applicable to CSHKP flares, and that they are affected by magnetic flux not being expelled from the current layer. However, it remains an ongoing task to comprehensively map out how 3D reconnection changes with parameters like plasma beta and guide field strength, and it is possible that reconnection could change between 3D plasmoid-mediated and Lazarian-Vishniac regimes as a solar flare progresses.

The properties of fluctuations in the reconnection layer are of significant interest for generation of Alfvénic waves. Spectral and anisotropy properties have been investigated by several authors. Huang and Bhattacharjee (2016) found elongated eddies aligned with the local magnetic field, scale independent anisotropy and an energy spectrum displaying an inertial range with $\tilde{E}_k \sim k^{-\alpha}$, $2.1 < \alpha < 2.5$. In a different setup (with periodic boundaries) and using different analysis techniques, Kowal et al. (2017) found somewhat different properties consistent with Kolmogorov-like turbulence and the Goldreich and Sridhar (1995, 1997) model, also highlighting that turbulence properties vary with the plasma beta.

The present situation is therefore that we have high confidence that self-generated turbulence is an intrinsic feature of 3D reconnection at high Lundquist numbers. In the CSHKP flare model (Figure 1), the leading edges of the flare ribbons are connected to the flare current sheet, so waves launched within the reconnection layer would be the first to be incident on the lower atmosphere. Simulations show that fluctuations in the reconnection layer have power law spectra, and solar observations support this too (French et al. 2021). However, work is ongoing to understand the properties of the turbulence and how they depend on the guide field and plasma beta. The theory of MHD turbulence is similarly in a period of active development, with ongoing efforts to understand the role of reconnection within turbulence (e.g., Loureiro and Boldyrev 2017; Dong et al. 2022).

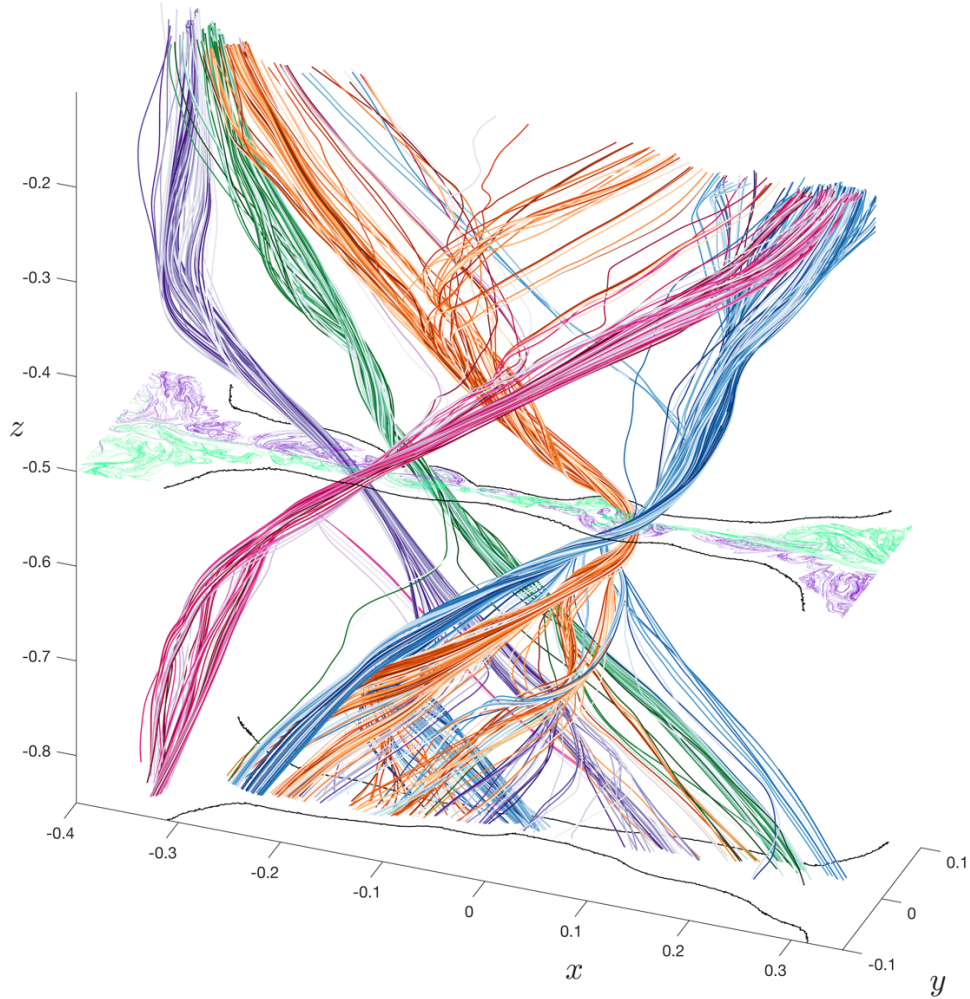

Figure 7 – Frayed flux ropes within 3D self-generated magnetic reconnection. Over short distances in *z*, magnetic field line reveal dynamically evolving twisted oblique flux ropes. Over longer distances, field lines become highly mixed. Flux rope dynamics, formation of localized twists and the magnetic stochasticity all have implications for wave generation inside the reconnection layer. Reproduced from Beg, Russell, and Hornig (2022) under a CC BY 4.0 license.

### 6.3 Braking of Reconnection Outflows

The CSHKP flare model includes a sunward reconnection outflow (**b** in Figure 1) that impinges on the closed magnetic loops below. The flow is therefore forced to brake in the above-the-loop region (**c** in Figure 1). There are various physical possibilities for how this might occur, but everyday experience of filling a sink or bath from the tap instructs us that conversion of bulk directed kinetic energy into turbulent (i.e., random small-scale) motions provides a natural and effective solution. The ideas that converting kinetic energy of the reconnection outflow into small scale vortical motions could be central to flare physics and explain looptop nonthermal broadening have existed for at least thirty years (Larosa and Moore 1993; Larosa, Moore, and Shore 1994; Tsuneta 1995; Haerendel 2009). However, simulations and observations have recently brought this mechanism to the fore.

We begin by deducing the properties of fluctuations produced by flow braking in magnetized plasma. Bending of magnetic field lines is opposed by the magnetic tension force, which is inversely proportional to the field line's radius of curvature. This force exerts a stabilizing influence on the development of small-scale motions in 2D planes containing the background magnetic field $\boldsymbol{B}$. However, the magnetic tension does not have this stabilizing effect on motions in planes perpendicular to $\boldsymbol{B}$. Applying these principles to the CSHKP model, turbulent braking requires dimensional freedom in the direction normal to the plane of Figure 1, hence we look for it only in flare simulations that have 3D freedom. We further anticipate that the small-scale motions produced by flow braking will have velocities virtually perpendicular to $\boldsymbol{B}$, and wavevectors satisfying $k_\parallel \ll k_\perp$ ($\boldsymbol{k} \cdot \boldsymbol{B} \approx 0$). Under these assumptions, magnetic perturbations are also perpendicular to $\boldsymbol{B}$, and evaluating the curls shows that vorticity $\boldsymbol{\omega}$ and current density $\boldsymbol{j}$ are virtually parallel (or antiparallel) to $\boldsymbol{B}$. Given the expected directions of velocity, vorticity and current density, these small-scale motions take the form of Alfvénic waves. Finally, while the waves should be anisotropic with $k_\parallel \ll k_\perp$, the parallel wavelength $\lambda_\parallel$ cannot be infinite due to finite extent of the wave braking region. The $\lambda_\parallel$ is likely to be determined by the width of the above-the-loop region, with the frequency determined by $f = v_A/\lambda_\parallel$.

The physical reasoning stated above has very recently received the support of MHD simulations that study the CSHKP flare model with 3D freedom and high resolution (Shen et al. 2022; Shen et al. 2023; Chen et al. 2020; Ruan, Yan, and Keppens 2023). These works have indeed found that flow braking produces small-scale vortical motions of an Alfvénic nature in the above-the-loop region. Furthermore, forward modelling these simulations produces symmetric gaussian lines that are non-thermally broadened to widths consistent with EIS and *IRIS* observations.

Figure 8 shows an example from Ruan, Yan, and Keppens (2023). The left panel displays the nonthermal width of a synthesized EIS Fe XXIV line, which reaches 200 km s$^{-1}$ near the apex in the above-the-loop region. Cyan contours represent AIA 131 Å intensity, showing the locations of flare loops. The phenomenology has much in common with the observation by Stores, Jeffrey, and Kontar (2021) shown in Figure 4.

The simulations also help to elaborate on the physics of how the small-scale motions are produced. Ruan, Yan, and Keppens (2023) highlighted that collision of sunward reconnection flows with the underlying arcade forms reflected upflows that interleave with downflows in the above-the-loop region, as shown in the right panel of Figure 8. This interleaving relies on the vertical component of the plasma velocity being able to reverse along the direction perpendicular to the plane of Figure 1, hence it was not captured by earlier 2.5D simulations. Ruan, Yan, and Keppens (2023) propose that the velocity shears between the downflows and reflected upflows feeds a Kelvin-Helmholtz instability that generates small-scale vortical motions. This process appears to dominate in the above-the-loop region where nonthermal velocities are greatest. Shen et al. (2023) computed power spectra in the above-the-loop region for their simulation, obtaining a one-dimensional kinetic energy spectrum proportional to $k^{-2.0}$.

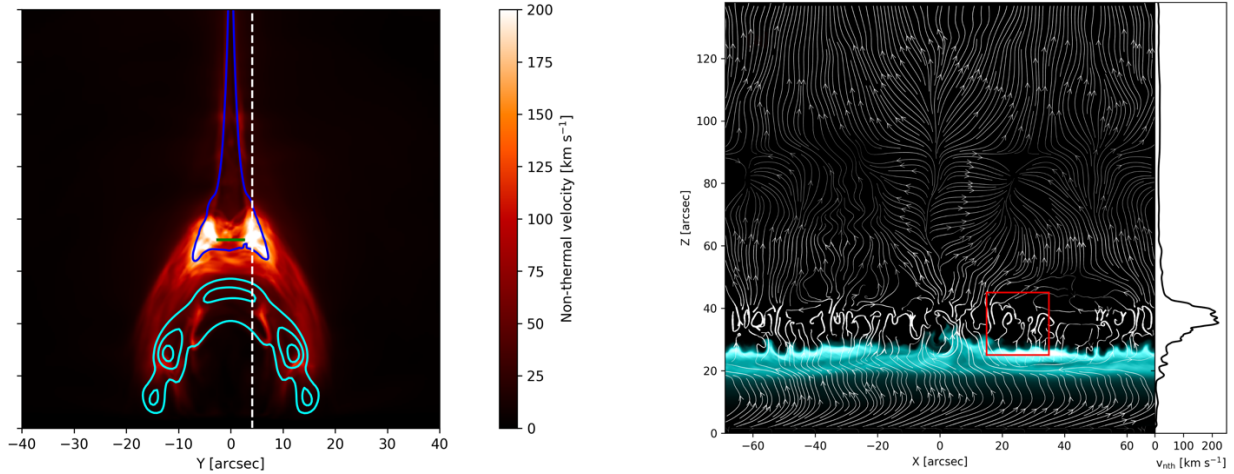

Figure 8 – MHD simulation demonstrating generation of small-scale motions by braking of the sunward reconnection flow. The left panel shows the nonthermal line widths of synthetic Fe XXIV profiles, with 131 Å intensity overlayed as cyan contours. The simulation produces nonthermal velocities of hundreds of km s$^{-1}$ that peak at the apex of the above-the-loop region, consistent with the observation shown in Figure 4. The right panel presents a side view, on a cross-section at the location indicated by the white dashed line on the left panel. Streamlines (white) show small-scale motions above the flare loops (cyan). The line plot at the righthand side shows $v_{nth}$, which peaks in the turbulent braking region. Reproduced from Ruan, Yan, and Keppens (2023) under a CC BY 4.0 license.

Globally, flow braking must be accomplished with force balance globally so that linear momentum is conserved. As pointed out by Haerendel (2009), the upward force that slows the bulk reconnection downflow at the looptops is ultimately balanced by an opposite force on the body of the Sun, requiring transmission of force between the loop apex and the lower atmosphere. Alfvénic waves are the natural agent. Thus, a comprehensive view of wave generation by flow braking encompasses not only loop-top turbulence, but also waves that are transmitted to the lower atmosphere. These include quasi-stationary waves that transmit the slowly-varying component of the force, in addition to which excitation of higher frequency propagating Alfvénic waves is also expected, given that reconnection outflows and their braking is highly variable in space and time. Ruan, Yan, and Keppens (2023) confirmed this viewpoint, reporting that fluctuations propagate along the magnetic field to the footpoints as Alfvén waves.

Generation of Alfvénic waves by flow braking is largely agnostic about the details of the reconnection process so long as reconnection is fast, despite those details being important for wave generation and particle acceleration within the reconnection layer itself. The simulation by Ruan, Yan, and Keppens (2023) used a nonuniform resistivity with a profile that typically produces Petschek reconnection, while Shen et al. (2022) considered two uniform resistivity cases on either side of the plasmoid instability threshold. For wave generation by flow braking, what matters is simply that there is a sunward reconnection flow to be stopped, and potentially the kinetic energy density of that flow.

# 7. TRANSPORT TO THE LOWER SOLAR ATMOSPHERE

## 7.1 General Remarks

We now examine the transport of energy to the lower solar atmosphere and the ensuing dynamics. Sections 7.2 and 7.3 describe the evolution of Alfvén waves in the corona, Section 7.4 addresses energy transmission across the transition region, and Section 7.5 considers the response of the lower atmosphere.

The earliest paper we are aware of that explored Alfvénic transport to the lower atmosphere in depth is by Emslie and Sturrock (1982). That pioneering work examined many of the points covered in this section, seeking a solution to the problem of temperature minimum region (TMR) heating during flares. The topic of flare-generated Alfvén waves was later explored by Melrose (1992) and Wheatland and Melrose (1994) with an emphasis on the propagation of Alfvénic fronts, e.g., created by switch-on of the energy release. The landmark paper of Fletcher and Hudson (2008) started a wider discussion of wave transport in solar flares, inspired in part by then recent microwave measurements of the coronal magnetic field strength and evidence that the photospheric magnetic field changes during flares. The paper is especially well known for highlighting the interesting possibility for Alfvénic waves/turbulence to accelerate particles locally in the loop legs and footpoints, which was further explored by Brown et al. (2009), McClements and Fletcher (2009) and Haerendel (2012). Important work by Haerendel (2009) also helped to define the modern version of the problem, proposing flow braking as a generator of Alfvénic waves, and raising the possibility that wave heating could be a driver of chromospheric evaporation.

## 7.2 Coronal Transport

We begin by examining how $v_{nth}$ is expected to vary in the corona, noting that observations like the nonthermal velocity map in Figure 4 show a decrease by about 50% from the apex to the footpoints.

It is instructive to first neglect reflection and dissipation to obtain a simple reference case. Under these assumptions, using the principle that the energy density of the wave is transported at the Alfvén speed, $\langle S_\parallel \rangle A$ remains constant in a Lagrangian frame moving with the wave, where $A$ is the cross-sectional area of the flux tube. Then, since $A = \Phi/B$ where $\Phi$ is the (constant) magnetic flux, and $\langle S_\parallel \rangle = v_A m_i n \langle v^2 \rangle$ as discussed in Section 5, the wave velocity amplitude is independent of $B$ and scales simply as $n^{-1/4}$. (The amplitude of the magnetic field perturbation meanwhile scales as $n^{1/4}$.) This well-known scaling result can also be derived using WKB analysis.

In Section 5, we considered typical values for $v_{nth}$ of 250 km s$^{-1}$ in the looptop and 120 km s$^{-1}$ in the footpoints (based on the observations described in Section 4). The resulting velocity ratio would be accounted for by the density alone if the footpoints were denser than the looptop by a factor 19. In a hydrostatic equilibrium, such a density ratio corresponds to just under three density scale heights. For comparison, the scale height $H = k_B T/m_i g$ for a 10 MK plasma is 300 Mm (0.44 $R_\odot$), so $3H$ is considerably taller than normal flare arcades at flare temperature. It is possible that such a density contrast might be produced by chromospheric evaporation, which is a dynamic effect; however, since nonthermal velocity maps appear to show a gradual decrease in $v_{nth}$ along the length of the loops (e.g. Figure 4) it appears likely that reflection and/or dissipation are significant.

Alfvén waves partially reflect if they have parallel wavelengths longer than the local length scale of the equilibrium (parallel to the magnetic field); otherwise, they propagate without significant reflection, as in a WKB solution. In this context we highlight the possibility of reflection being enhanced by small-scale density variations, which have previously been invoked for the Sun's open corona (van Ballegooijen and Asgari-Targhi 2016). A hypothesis of reflection from small-scale density variations is difficult to falsify with current observations, so we leave this as an open possibility.

Waves propagating through a plasma can be dissipated by resistivity, viscosity or particle acceleration. The corona is typically regarded as a low-dissipation plasma, however, the properties of flaring plasma are markedly different to normal coronal properties, and in flares there is the possibility for short wavelengths and high frequencies, so we will consider these mechanisms explicitly.

Starting with resistive dissipation, linear MHD theory predicts that the Alfvén waves damp resistively with a characteristic damping length

$$L_{res} = \frac{v_A}{\eta k^2}$$

for the energy density, where we have used $\eta = 1/\mu_0 \sigma$ to mean the magnetic diffusivity and are using S.I. units. Different parallel and perpendicular diffusivities can also be incorporated, as shown in Section 7.5. Using the Spitzer conductivity (Spitzer 1962), a convenient formula for the magnetic diffusivity for coronal applications is

$$\eta = 1.0 \left(\frac{\ln \Lambda}{20}\right) \left(\frac{T}{10^6 \text{ K}}\right)^{-3/2} \text{ m}^2 \text{ s}^{-1}.$$

Setting the Coulomb logarithm to 20 and the flare temperature to 10 MK yields a magnetic diffusivity of 0.032 m² s⁻¹. Subsequently, obtaining $L_{res}$ of 50 Mm or less for an Alfvén speed of 2800 km s⁻¹ requires wavelengths below 4.7 m. We conclude that resistivity does not significantly damp large-scale Alfvén waves in the corona, but it may provide dissipation for the shortest scale waves produced during flares, such as those reached by turbulent cascade.

For context, the ion kinetic scales (e.g., Shibata and Tanuma 2001) are the ion inertial length,

$$\delta_i = 3.0 \left(\frac{n}{10^{16} \text{ m}^{-3}}\right)^{-1/2} \text{ m},$$

and the ion Larmor radius,

$$r_{L,i} = 0.1 \left(\frac{B}{100 \text{ G}}\right)^{-1} \left(\frac{T}{10^6 \text{ K}}\right)^{1/2} \text{ m}.$$

For a number density of 10¹⁷ m⁻³, $\delta_i$ is about 1 m, whereas for a flare magnetic field strength of 400 G and temperature of 10 MK, $r_{L,i}$ is about 8 cm. Comparing these values to $L_{res}$ confirms that resistive damping could be a significant for the shortest wavelengths and therefore affect the dissipative range of any turbulent cascade, although collisionless effects due to $\delta_i$ also become significant at similar scales.

Viscous damping of Alfvén waves is more complicated than resistive damping because the Braginskii viscous stress tensor is highly anisotropic in magnetized plasma. In this context, we will use $\eta$ with a numerical subscript to represent the viscosity coefficients, which are given by

$$\eta_0 = 1.1 \times 10^{-2} \left(\frac{\ln \Lambda}{20}\right) \left(\frac{T}{10^6 \text{ K}}\right)^{5/2} \text{ kg m}^{-1} \text{ s}^{-1},$$

$$4\eta_1 = \eta_2 = 1.25 (\Omega_i \tau_i)^{-2} \eta_0,$$

where $\Omega_i = eB/m_i$ is the proton gyrofrequency and

$$\tau_i = 7.5 \times 10^{-2} \left(\frac{T}{10^6 \text{ K}}\right)^{3/2} \left(\frac{n}{10^{16} \text{ m}^{-3}}\right)^{-1} \text{ s}$$

is the proton collision time.

For our usual above-the-loop parameters, $\tau_i$ is 0.24 s and $\Omega_i$ is $3.8 \times 10^6$ Hz, so the plasma is strongly magnetized, with $\Omega_i \tau_i \sim 10^6 \gg 1$. Then, setting the Coulomb logarithm to 20, we estimate $\eta_0 = 3.5$ kg m$^{-1}$ s$^{-1}$, and obtain $\eta_1 = 1.3 \times 10^{-12}$ kg m$^{-1}$ s$^{-1}$ with $\eta_2 = 4\eta_1$.

In the linear theory, e.g., Equation (8.40) of Braginskii (1965),

$$L_{visc}^{linear} = \frac{v_A \rho_0}{\eta_1 k_\perp^2 + \eta_2 k_\parallel^2}.$$

Considering cases with $k_\perp^2 \gg k_\parallel^2$, obtaining $L_{visc}^{linear}$ of 50 Mm or less for an Alfvén speed of 2800 km s$^{-1}$ requires wavelengths of 2.8 m or less. We also remark that the temperature scaling of $\eta_1$ and $\eta_2$ is $\sim T^{-1/2}$ compared to $\sim T^{-3/2}$ for the resistivity, so viscous dissipation reduces more slowly with temperature than resistive dissipation does. These results show that linear viscous dissipation becomes effective in flares at similar scales to resistive dissipation and ion inertial effects.

The vast ratio of the parallel to perpendicular viscosity coefficients means that nonlinear viscous damping via the $\eta_0$ viscosity coefficient should also be considered for coronal Alfvén waves (Nocera, Priest, and Hollweg 1986; Russell 2023). Nonlinear wave damping is potentially more important in flaring plasma than it is in the quiet corona, because the ratio of viscosity coefficients increases nonlinearly with temperature, as $\eta_0/\eta_1 \sim T^3$. Thus, the ratio of the viscosity coefficients is a factor 1000 greater for 10 MK plasma than it is for 1 MK plasma, significantly increasing the relative effectiveness of damping via the $\eta_0$ viscosity coefficient. The nonlinear damping also increases in absolute terms, since $\eta_0 \sim T^{5/2}$.

The nonlinear decay length for linearly polarized shear Alfvén waves due to parallel wavelengths and parallel viscosity (Russell 2023) can be expressed as

$$L_{visc}^{nonlinear} = 1.2 \times 10^{12} \left(\frac{\ln \Lambda}{20}\right) \left(\frac{f}{1 \text{ Hz}}\right)^{-2} \left(\frac{T}{10^6 \text{ K}}\right)^{-5/2} \left(\frac{B}{100 \text{ G}}\right)^{5} \left(\frac{n}{10^{16} \text{ m}^{-3}}\right)^{-3/2} \left(\frac{\langle v^2 \rangle^{\frac{1}{2}}}{10^5 \text{ m s}^{-1}}\right)^{-2} \text{ m},$$

where we have dropped the finite-beta correction since a 10 MK plasma has a sound speed of 370 km s$^{-1}$ compared to an assumed Alfvén speed of 2800 km s$^{-1}$, giving $(c_s/v_A)^2 = 0.017$. Considering our usual above-the-loop values for $\ln \Lambda$, $B$, $T$ and $n$, and an rms velocity of 250 km s$^{-1}$, this expression reduces to

$$L_{visc}^{nonlinear} = 1.9 \times 10^{10} \left(\frac{f}{1 \text{ Hz}}\right)^{-2} \text{ m}.$$

Obtaining $L_{visc}^{nonlinear}$ of 50 Mm or less from this result requires wave frequencies of 20 Hz or greater. We can therefore neglect nonlinear viscous damping for frequencies of around 1 Hz for our chosen parameters, but nonlinear viscous damping appears potentially significant for the higher-frequency part of the wave spectrum. We also remark that while nonlinear viscous damping is not significant at lower frequencies for our chosen parameters, a broader parameter survey may yet identify conditions for which it is.

The appearance of nonlinear viscous damping in hot tenuous plasma is associated with low collisionality, and collisionless effects may become important for higher frequency waves. The traditional derivation of Braginskii MHD requires $k_\parallel \lambda_{mfp} < 1$, where $\lambda_{mfp} = v_{Ti}\tau_i$ is the ion mean-free path along the magnetic field with ion thermal speed $v_{Ti} = \sqrt{k_B T/m_i}$. This corresponds to the Reynolds number (based on $\eta_0$, $l = k_\parallel^{-1}$ and the ion sound speed) being greater than unity (Section 6.3 of Russell (2023)). Alternatively, it can be expressed as the following condition on the frequency,

$$f < f_c = \frac{v_A/v_{Ti}}{2\pi\tau_i} = 51 \left(\frac{T}{10^6 \text{ K}}\right)^{-2} \left(\frac{B}{100 \text{ G}}\right) \left(\frac{n}{10^{16} \text{ m}^{-3}}\right)^{1/2}.$$

The parameter values we considered for the above-the-loop region yield $f_c = 7$ Hz. Considering frequencies lower than $f_c$ returns $L_{visc}^{nonlinear}$ of 390 Mm or greater, which is too large to matter for most flares. Damping of higher frequency waves with $f > f_c$ may be collisionless in nature. For example, Squire, Quataert, and Schekochihin (2016) have shown that linearly polarized shear Alfvén waves with amplitudes $\delta B/B_0$ exceeding a threshold $\beta^{-1/2}$ are interrupted by parallel firehose instability.

The last coronal dissipation mechanism we consider is particle acceleration. This is a particularly promising avenue, considering the abundant evidence that solar flares are efficient particle accelerators. This is especially true in the above-the-loop region where nonthermal velocities are greatest and a very large fraction of the electrons can be nonthermal (see discussion and references in Section 2). Kontar et al. (2017) used joint *RHESSI*-EIS observations to infer that the time scale for Alfvénic energy density to energize electrons is 1 to 10 s, which is consistent with the turbulence time scale for Goldreich-Shridhar type turbulence (see discussion in Kontar et al. (2017)). On the theory side, the idea that Alfvénic waves and/or turbulence accelerate particles to high energies (Fermi 1949; Parker 1957b) is only slightly younger than the famously long-standing coronal heating problem, but uncertainty over the details of particle acceleration by waves has endured. This topic is rightly the subject of separate reviews, including Miller et al. (1997), Zharkova et al. (2011) and Petrosian (2012). We also highlight papers by Fletcher and Hudson (2008), Brown et al. (2009), McClements and Fletcher (2009), Bian, Kontar, and Brown (2010) and Haerendel (2012).

In summary, nonthermal broadening measurements indicate that the velocity amplitude of Alfvénic waves decreases by about 50% from the apex to the footpoint. This is unlikely to be accounted for by simple WKB density scaling and is likely to require either reflection (e.g., from small-scale inhomogeneities) or dissipation. Linear resistive and viscous dissipation both require short length scales (e.g., perpendicular wavelengths) of several meters. Thus, these processes do not damp large-scale Alfvénic waves, but they may be important at the conclusion of a turbulent cascade. Nonlinear viscous damping appears unlikely to damp Alfvén waves in flares generally, although it may damp the high frequency component of the wave spectrum. Collisionless processes may be important, and we note that when collisions are scarce Alfvén waves can develop significant pressure anisotropy, triggering instabilities such as firehose instability. Finally, given that solar flares are powerful particle accelerators, the most obvious answer to how the waves damp may be that their energy is transferred to nonthermal particles, although the details of particle acceleration by waves/turbulence remains an active research subject.

### 7.3 Magnetic Convergence and Phase Mixing

Having explored the processes governing wave amplitude, we now explore how transport affects the perpendicular wavelength. Here, we consider concentration of wave energy as the magnetic field converges towards the footpoints, as well as several types of phase-mixing, which describes the shortening of perpendicular wavelengths as wave fronts are distorted during propagation along field lines.

The simplest effect that modifies perpendicular wavelengths is the convergence (or divergence) of field lines: if field line spacing becomes closer, then maintaining the same number of phase cycles implies the transverse wavelength is shorter. If convergence occurs in one transverse direction, $k_\perp \sim B$; if it occurs in both transverse directions, $k_\perp \sim B^{1/2}$.

The basic form of phase mixing (Heyvaerts and Priest 1983) occurs due to transverse gradients of $v_A$, for example due to density gradients. The corona has a fine-stranded structure, with the strands most frequently having full widths of 500–550 km (Brooks et al. 2013; Aschwanden and Peter 2017; Williams et al. 2020). These values are comparable to 310 km elemental widths inferred from observations of coronal rain by Antolin and Rouppe van der Voort (2012); while widths for other types of fibrillar structure in the Sun are similar but sometimes narrower, often around 100 km. Accordingly, basic phase-mixing implies that Alfvénic waves incident on the transition region should have perpendicular wavelengths not more than several hundred km after propagating through coronal fine structure.

Phase mixing can also occur due to the magnetic field line geometry. Starting with the case of a smooth laminar field, neighboring field lines have different arc lengths from their apex to the lower boundary. Thus, if waves travel at the same speed along both field lines, then the part of the wave front that travels on the shorter field line reaches the boundary first. These time differences distort the wavefront, shortening the perpendicular wavelength. This geometrical phase-mixing effect has been treated analytically by Weinberg (1962) and applied to solar flare energy transport by Emslie and Sturrock (1982). It has also been demonstrated in a simulation by Russell and Stackhouse (2013). However, this form of phase mixing is likely the weakest of the three.

The third type of phase-mixing operates in non-laminar magnetic fields. In stochastic magnetic fields, the separation distance between two field lines that are initially neighbors increases exponentially with respect to distance along the field. In this scenario, field-aligned transport produces an exponential shortening of perpendicular wavelengths, which accesses small scales at an exponential rate (Similon and Sudan 1989). Considering the role of magnetic stochasticity in various models of fast magnetic reconnection (Section 6.2) and the possibility for reconnection in flow-braking turbulence (Section 6.3) to create magnetic stochasticity, it appears highly plausible that this type of phase mixing will apply to Alfvénic waves in solar flares. Future work with MHD flare simulations will be valuable for identifying the degree of stochasticity produced in flares and its impact on perpendicular wavelengths.

The effects described above may provide a first stage in shortening the transverse wavelength. As waves develop larger $k_\perp$, the strengthened velocity and magnetic shears can trigger Kelvin-Helmholtz (KH) and tearing instabilities, which may accelerate the cross-scale transport. Heyvaerts and Priest (1983) included this possibility in their landmark paper on phase mixing, concluding that propagating wave are KH stable but standing waves are KH unstable. Applied to flare transport, one might therefore anticipate instabilities of phase mixed waves playing a role where waves reflect, e.g., from the transition region. Haerendel (2009) proposed a two-stage mechanism of this type could operate in flares, and it is well publicized that KH instability of phase mixed waves hastens dissipation in the boundaries of kink oscillating coronal loops (Terradas et al. 2008; Antolin et al. 2017). Turbulent cascade may also play a role (e.g., Howes and Nielson 2013, and references therein).

## 7.4 Energy Transmission to the Chromosphere

The transition region is normally marked by a change in temperature over a relatively short distance, from around 10,000 K in the chromosphere to about 1 MK in the corona (here we consider a preflare coronal temperature). Pressure balance implies that the Alfvén speed changes on the same scale. For example, assuming a neutral-dominated chromosphere, pressure balance yields $2n_{cor}T_{cor} = n_{chr}T_{chr}$, where $n_{cor}$ is a proton density and $n_{chr}$ is a the total density of protons and hydrogen atoms, so that the total mass density increases by a factor of approximately 200. This is turn decreases the Alfvén speed by a factor 14, where the chromospheric $v_A$ is calculated using the total mass density because neutrals are collisionally coupled to the ions.

Alfvénic waves with parallel wavelengths much longer than the depth of the transition region see it as a step in the wave speed. The energy transmission coefficient (Emslie and Sturrock 1982; Fletcher and Hudson 2008; Russell and Fletcher 2013) is

$$\mathcal{T} = \frac{4\theta^{1/2}}{(1 + \theta^{1/2})^2},$$

where $\theta = (v_{A,chr}/v_{A,cor})^2 \approx 0.5\, T_{chr}/T_{cor}$. Thus, $\mathcal{T} \approx 2\sqrt{2}\sqrt{T_{chr}/T_{cor}} \approx 0.28$ for the initial temperatures discussed above, meaning that 28% of the wave energy crosses the transition region per incidence. This permits significant transmission of wave energy to the lower atmosphere, from the first times waves are generated in the corona. For shorter wavelengths, transmission exceeds this lower bound, approaching 100% transmission for the shortest wavelengths (Russell and Fletcher 2013; Reep et al. 2018).

Alfvén waves have $k_\parallel = \omega/v_A$ so the parallel wavelength scales as $\lambda_\parallel \sim v_A$. The parallel wavelength is therefore shortened considerably (factor about 14) when the wave is transmitted to the chromosphere. If the new wavelength is shorter than the scale height for the Alfvén wave in the chromosphere, it will propagate there without further reflection. Otherwise, wave energy will continue to be reflected. Putting the whole system together, the percentage of incident energy retained in the chromosphere increases with frequency. Numerical studies by Russell and Fletcher (2013) (their Figure 5) and Reep et al. (2018) (their Figure 3) have investigated the functional dependence using semi-empirical models of the lower atmosphere. As a rule of thumb, energy transmission for frequencies above 1 Hz exceed 20–50% per incidence, depending on the atmospheric model.

## 7.5    Chromospheric Heating

The chromospheric dissipation of downgoing waves was first investigated by Emslie and Sturrock (1982), using a WKB solution of the linearized single-fluid MHD equations. Their Figure 2 neatly summarizes how the heating rate peaks at different heights for different wave parameters. The heating rate for waves with 1 Hz frequencies and transverse wavelengths of 1–10 km or greater peaks in the temperature minimum region (TMR), and Emslie and Sturrock (1982) proposed that this energy deposition might account for the TMR warming by 100–200 K in flares (Machado, Emslie, and Brown 1978). Figure 2 of Emslie and Sturrock (1982) also shows that waves with higher frequencies and/or shorter perpendicular wavelengths are stopped higher in the chromosphere. While this aspect was not developed further by Emslie and Sturrock (1982), wave heating at these higher altitudes causes a much greater temperature increase on account of the lower density, and it has subsequently been proposed as a driver of chromospheric evaporation, joining energy deposition by thermal conduction and particle beams (Haerendel 2009; Reep and Russell 2016).

One of the major dissipation processes in the chromosphere is ion-neutral friction. At chromospheric temperatures, protons and neutral hydrogen are coupled by resonant charge exchange, which operates with a momentum transfer collision frequency

$$\nu_{H,p} = 1.18 \times 10^4 \left(\frac{n_p}{10^{18} \text{ m}^{-3}}\right)\left(\frac{T}{10^4 \text{ K}}\right)^{1/2}\left(1 - 0.125\log\left(\frac{T}{10^4 \text{ K}}\right)\right)^2 \text{ s}^{-1}.$$

This formula is consistent with Equation (A10) of De Pontieu, Martens, and Hudson (2001), and Equation (23) of Russell and Fletcher (2013) after correcting the typo pointed out in Reep and Russell (2016). At the TMR, the hydrogen ionization degree becomes so low that singly-ionized metal ions dominate and the collision frequency is instead calculated from

$$\nu_{H,M^+} = 2.1 \times 10^3 \left(1 + \frac{1}{A_M}\right)^{-1/2}\left(\frac{n_{M^+}}{10^{18} \text{ m}^{-3}}\right) \text{ s}^{-1}$$

(De Pontieu, Martens, and Hudson 2001; Russell and Fletcher 2013). For chromospheric densities and temperatures, the ions and neutrals are strongly coupled with millisecond momentum transfer timescales. Thus, while neutral particles are not directly tied to the magnetic field by Lorentz force, they are tied to it indirectly by collisions with ions. (This is a very different parameter regime to the Earth's ionosphere, in which the coupling timescale is closer to hours, allowing existence of Hall and Pedersen currents.) While the collisional coupling in the chromosphere is very strong, it is imperfect, which leads to heating.

Resistive damping can also play a significant role in the chromosphere. Compared to the corona, the lower temperature enhances resistive damping by a factor greater than 1000. Resistive dissipation is also more effective in the chromosphere because $k_\parallel = \omega/v_A \sim n^{1/2}$ is larger due to the higher density, $k_\perp$ is likely to be larger due to additional phase mixing, and collisions with neutrals enhance the resistivity.

The effects of electron resistivity and ion-neutral friction can be combined using anisotropic Cowling resistivity. Here, we present a simplified version valid for low ionization degree and small $\omega/\nu_{ni}$, and refer readers to Reep and Russell (2016) for more general equations. The parallel diffusivity is

$$\eta_\parallel = \frac{m_e}{\mu_0 n_e e^2}(\nu_{ei} + \nu_{en}),$$

while the perpendicular diffusivity is

$$\eta_\perp = \eta_\parallel + \eta_C, \quad \eta_C = \frac{B^2 \rho_n}{\mu_0 \nu_{ni} \rho_t^2}.$$

The wave energy damping length (Reep and Russell 2016) is then

$$L_D = \frac{v_A}{\eta_\parallel k_\perp^2 + \eta_\perp k_\parallel^2} = \frac{v_A}{\eta_\parallel k^2 + \eta_C k_\parallel^2}.$$

This formula implies that damping by ion-neutral friction occurs through $k_\parallel = \omega/v_A$, whereas damping by electron resistivity is determined by $k$.

Russell and Fletcher (2013) modelled wave propagation from the corona to the chromosphere, and the ensuing damping, modelling the neutrals and plasma as separate fluids that are collisional coupled. For simplicity, that work solved the linearized equations in the case $k_\perp = 0$. The results showed that a 1 s wave pulse, for example, deposits most of its energy at the TMR by ion neutral friction, but the greatest temperature increases occur in the upper chromosphere.

The full problem includes nonlinear feedbacks since heating the atmosphere causes it to evolve. The ensuing evolution has been investigated by adding wave heating models to the HYDRAD and RADYN 1D radiative hydrodynamic codes (Reep and Russell 2016; Kerr et al. 2016; Reep et al. 2018). Figure 9 shows HYDRAD results from Reep and Russell (2016) that compare the atmospheric response for two cases of Alfvénic wave heating (top two rows) and one case of electron beam heating (bottom row). These results demonstrate that wave heating can produce gentle or explosive evaporations with upflows at several hundred km s$^{-1}$, resembling those produced by electron beam heating.

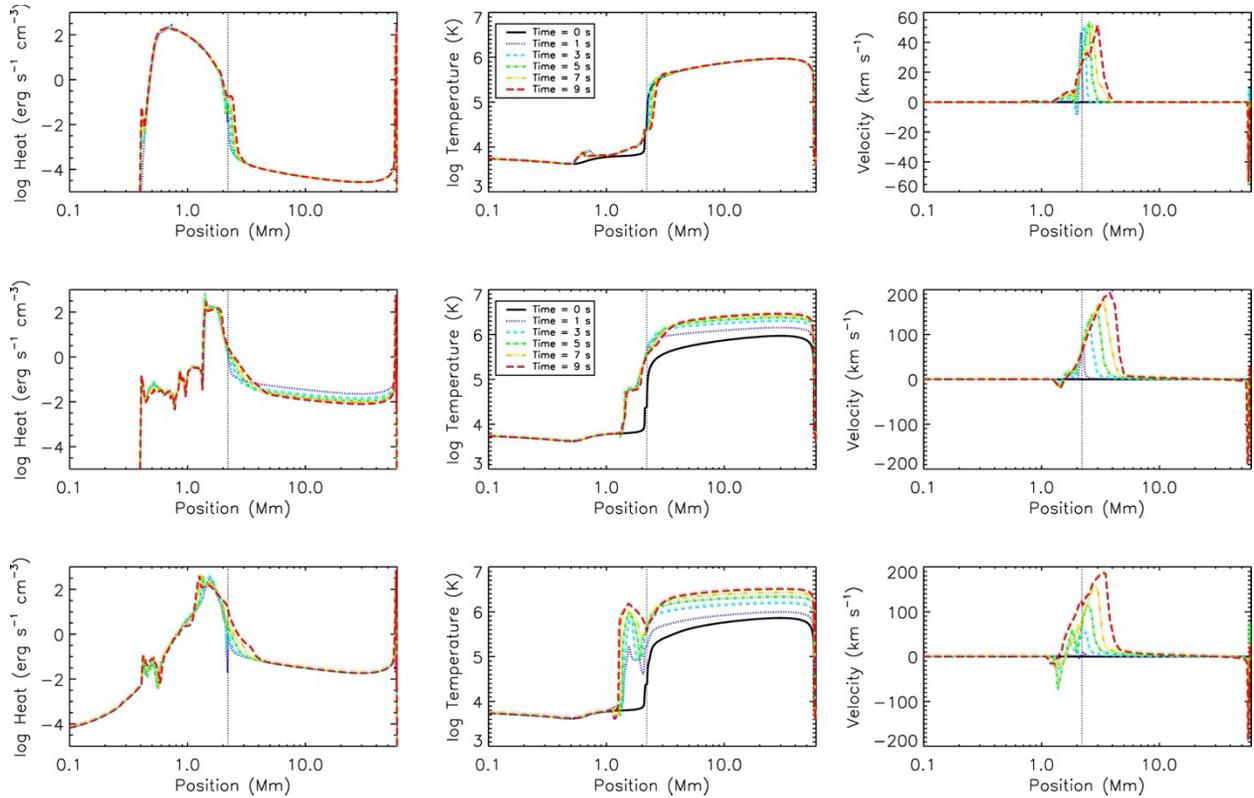

Figure 9 – Evolution of model solar atmospheres in three scenarios. The top row shows the response to waves injected with $f = 10\ Hz$ and $k_\perp = 10^{-3}\ m^{-1}$, the middle row to waves injected with $f = 10\ Hz$ and $k_\perp = 4 \times 10^{-2}\ m^{-1}$, and the bottom row to an electron beam with low-energy cutoff of 20 keV and spectral index $\delta = 5$. All simulations have an energy flux of $10^7$ W m$^{-2}$, consistent with estimates in Section 5. The upflows (right column) for the middle and bottom rows are virtually indistinguishable. Reproduced from Reep and Russell (2016). © AAS. Reproduced with permission.

The significant caveat is that producing chromospheric evaporation requires relatively high frequencies and/or short perpendicular scales below 10 km, consistent with expectations from Emslie and Sturrock (1982), Haerendel (2009) and Russell and Fletcher (2013). It is currently unclear what fraction of wave energy produced by reconnection and flow braking enters the lower atmosphere with the properties needed to damp strongly in the upper atmosphere. The stochastic phase mixing (Similon and Sudan 1989) discussed in Section 7.3 offers a powerful mechanism for transferring energy to short perpendicular wavelengths, as do the mechanisms of phase mixing combined with instabilities or turbulent cascade.

The general similarity of the upflows produced by short wavelength Alfvénic waves and electron beams was further confirmed using RADYN by Kerr et al. (2016), who also synthesized various chromospheric spectral lines. An intriguing finding was that that Mg II profiles for their wave heated simulation were different to those for the electron beam simulation, with the wave heating simulation better matching *IRIS* observations. This finding suggests that careful study of lines formed in the mid to upper chromosphere may be capable of distinguishing between different heating mechanisms.

Further differences have been identified by Reep et al. (2018) who used a second-generation wave driver that replaced the assumption of instantaneous transport with a new method that captures the effects of finite wave travel time. In principle, the fact that Alfvénic waves take time to propagate (whereas electron beams travel at close to the speed of light) can lead to different dynamics and observational signatures. Reep et al. (2018) especially highlighted that as waves heat a particular layer in the chromosphere, the temperature and ionization fraction increase, decreasing the parallel and Cowling resistivities in that layer. Subsequent waves therefore penetrate deeper into the atmosphere before damping, such that waves "bore a hole" deeper into the chromosphere. This behavior is opposite to an electron beam, which heats at progressively greater heights as chromospheric evaporation raises the coronal density (assuming the beam properties remain constant). As a result, differences develop in the ionization, velocity and pressure profiles that might lead to further opportunities to observationally distinguish heating by Alfvénic waves from heating by electron beams or thermal conduction.

## 8. FUTURE OUTLOOK

In the last 15 years, the topic of Alfvén waves in solar flares has matured significantly and several game-changing developments have given us high expectations for coming years. Here, we identify four major shifts that are anticipated or already happening, and the scientific questions they will help address.

### 8.1 High-performance 3D Simulations

We have recently been fortunate to see 3D simulations achieve resolutions that transform the landscape for exploring the role of turbulence in solar flares. In particular, we highlighted in Section 6.2 how advanced simulations opened up the investigation of 3D magnetic reconnection at high Lundquist numbers, and in Section 6.3 how they have begun illuminating the generation of Alfvénic motions by flow braking in the above-the-loop region.

We anticipate that the new generation of 3D simulations will address the generation mechanisms for Alfvénic waves/turbulence in solar flares, establish the resulting wave properties and connect to observations. Many things need to happen to fully grasp the opportunity; but much can be achieved by sustaining the development of 3D CSHKP flare simulations and allowing Moore's law to reveal a wider inertial range of the spectra. We are also hopeful that these topics will be approached not only with MHD, but also incorporating kinetic-type techniques to better model the reconnection process, and to self-consistently model particle acceleration. We believe this should be a high priority: done in coordination with theory and observations, the potential prizes include solving magnetic reconnection and the major workings of solar flares.

### 8.2 MUSE

Successive generations of spectroscopic observations have been the crucial source of knowledge for the amplitudes of small-scale plasma motions in solar flares, with hot lines especially important. The major axes of improvement for the instrument design have long been spatial resolution, spectral resolution, cadence and instrument sensitivity. We have also seen, with *Hinode*/EIS and *IRIS* especially, the value of imaging, even with the low effective cadence that comes from rastering a single slit across a flare.

A wishlist for a new instrument to study Alfvénic waves in solar flares would include: coverage of hot lines to capture the plasma with the largest nonthermal velocities; coverage of non-flare lines in the corona and transition region to investigate motions in these layers; spatial resolution comparable to or better than *IRIS* to spatially resolve footpoint emission; the best possible sensitivity to measure velocities as early as possible in the flare; short exposure times; and ability to produce images with rapid cadence.

The NASA-selected *Multi-slit Solar Explorer* (*MUSE*) mission (De Pontieu et al. 2020) delivers in many of these respects. *MUSE* will carry a multi-slit spectrograph (*MUSE*/SG) and a high-resolution EUV contextual imager (*MUSE*/CI). *MUSE*/SG builds on the heritage of single-slit EUV spectrographs such as EIS and *IRIS*, but it is designed to obtain spectra from 37 slits simultaneously for bright lines of Fe IX 171 Å, Fe XV 284 Å, Fe XIX 108Å and Fe XXI 108 Å, which are formed at temperatures around 0.8 MK, 2.5 MK, 10 MK and 13 MK. Its design aims for spatial resolution of 0.4″ (about 400 km on the Sun), slit spacing of 4.45″, and a 170″ × 170″ field of view (FOV). Combined with a cadence of 1 s or less, it will be possible to either stare at many fixed slit locations simultaneously, providing comprehensively sampled measurements of $v_{nth}$ at timescales relevant to MHD turbulence and particle acceleration, or to raster the whole FOV in 12 s or less. The EUV imager *MUSE*/CI, meanwhile, will provide 304 Å and 195 Å images at 0.33″ resolution (for comparison the resolution of *SDO*/AIA (Lemen et al. 2012) is 1.5″) with a 580″ × 290″ FOV and cadence of 4 s for single channel operation (8 s for dual channel operation).

Science objectives for *MUSE* related to flares and eruptions have been discussed by (Cheung et al. 2022), and their Section 5.4 is especially relevant to the current article. We believe that *MUSE* has excellent potential to deliver a step change in observations of waves and turbulence in solar flares by measuring spectral properties with unrivalled spatial coverage and cadence. We also highlight that the combination of *MUSE*/SG and *MUSE*/CI is perfect for investigating motions in 3D. *MUSE* is currently estimated for launch in 2027, a timeframe that gives impetus to advancing the theory and modelling in preparation.

## 8.3 DKIST

New observational opportunities also include the new generation of ground-based solar telescopes. At the current state of the art, the Daniel K. Inouye Solar Telescope (DKIST; Rimmele et al. 2020) is equipped with a 4 m aperture and a high-order adaptive optics system, which are designed to resolve scales as short as 0.02″, or 20 km on the Sun. Its initial suite of five instruments, which includes four polarimeters, allow observations at visible and near-infrared wavelengths, from 380 to 5000 nm. First light was achieved in December 2019, and it has more recently started collecting scientific data. Other 4 m class solar telescopes are in development, including the European Solar Telescope (EST), which in combination with DKIST would provide near continuous solar observing.

Advancing understanding of solar flares and CMEs is a major science driver behind these telescopes, and Section 4 of the DKIST Critical Science Plan (Rast et al. 2021) details how DKIST is expected to contribute to flare science. Here, we briefly outline some of the major opportunities relevant to Alfvénic waves in solar flares.

A major objective for DKIST is to measure coronal magnetic fields, using principles demonstrated by (Lin, Penn, and Tomczyk 2000; Lin, Kuhn, and Coulter 2004) using a 0.46 m aperture telescope. The much larger DKIST has targets of "spatial resolution of one to two arcsec, sensitivity of order 10 G or less… with a temporal resolution of tens of seconds to minutes" (Rimmele et al. 2020). Such measurements would provide valuable contextual information for limb flares, complementing existing approaches of deducing magnetic field information from microwave observations (e.g., Chen et al. 2020;

Fleishman et al. 2022) and intensity images. Routine measurements of coronal magnetic field strength would be a very welcome achievement for constraining the Poynting fluxes in flares (see Section 5).

Ground-based polarimetry can also identify regions of strong turbulence and/or wave activity. As discussed in Section 4.7, French et al. (2019) showed that the linear polarization (Stokes $P$) fell to low values in the flare plasma sheet and the above-the-loop region of the X8.2 flare on 10 September 2017. The reduction at these locations was attributed to the orientation of the magnetic field varying on unresolved scales and/or along the line of sight. Stokes $P$ is therefore complementary to the nonthermal broadening explored in Section 4: nonthermal broadening indicates unresolved motions, whereas low $P$ can identify unresolved variations in the magnetic field orientation. When the two occur together, the case for interpreting the unresolved motions as Alfvénic waves or turbulence is strengthened, since a model of field-aligned flows does not explain the reduction in $P$.

Another major objective of DKIST is to obtain magnetic field data at multiple heights in the Sun's lower atmosphere, allowing scientists to probe the 3D magnetic structure. As discussed in the Introduction, solar flares and eruptions are powered by reconfiguration of the magnetic field, which produces rapid changes to the magnetic field in the lower atmosphere that are likely communicated by large-amplitude Alfvénic wave fronts. DKIST offers a good prospect of probing these magnetic field changes, and their association with particle acceleration (Fletcher and Hudson 2008), sunquakes (Hudson, Fisher, and Welsch 2008; Russell et al. 2016) and radiative emissions from the lower atmosphere (Johnstone, Petrie, and Sudol 2012; Russell and Fletcher 2013).

Imaging and spectroscopy of flare ribbons at the high cadence and high resolution provided by DKIST will help to explore energy transport in flares. As discussed in Section 7.5, 1D radiative hydrodynamics simulations show that the lower atmosphere responds differently to heating by Alfvénic waves than it does to heating by electron beams, although there are also some major similarities (Reep and Russell 2016; Kerr et al. 2016; Reep et al. 2018). DKIST data are especially important to constrain the size (Krucker et al. 2011) and height (Martínez Oliveros et al. 2012) of the white light sources.

Finally, since flare ribbons are believed to be connected to the coronal reconnection layer (Figure 1), their morphology and dynamics shed light on many of the central issues concerning Alfvénic waves in flares. For example, French et al. (2021) used images from the *IRIS* Slit Jaw Imager to track the exponential growth of power at different wavenumbers along the ribbons of a confined B-class flare. Instability began at a "key spatial scale" (1.75 Mm in the east ribbon), spread to smaller and larger scales and saturated with a $k^{-2.3}$ power law across all scales. Nonthermal broadening in the ribbons rose as the instability spread to all scales. The timeline and power law slope are consistent with the models of broadband Alfvénic wave generation within the reconnection layer discussed in Sections 6.1 and 6.2. Higher resolution studies using DKIST, covering more events, are therefore an exciting prospect to advance understanding of the wave generation and the wave spectrum.

### 8.4 Integrated Modelling

There is valuable scientific work to be done advancing the subtopics covered in this article individually, and integrating the component parts into the whole system. As computing power advances further, we look forward to a combined treatment of the fast reconnection process and outflow braking in CHSKP flare geometries in the relatively near future. Thence, modelling approaches can be brought to bear that have already been developed through other work, including connections with the lower atmosphere, particle energization, radiative effects, and forward modelling to compare with observations (e.g., Cheung et al. 2019; Gordovskyy et al. 2020; Ruan, Xia, and Keppens 2020). The end goal is to model flares from reconnection to radiation, capturing the energy pathways in between.

We also highlight the role for simulations tailored to specific events. 3D data-driven simulations (e.g., Amari, Canou, and Aly 2014) have greatly advanced understanding of the mechanisms responsible for triggering solar eruptions and flares. The necessity that such simulations model an entire active region and its environment means the shortest resolved scale is significantly larger than for more idealized simulations. Hence, we must wait before wave excitation processes are captured in full data-driven simulations. Once this occurs, however, the ability to directly compare synthetic and real observations will be a great aid to refine our knowledge of Alfvénic waves in flares.

## 9. SUMMARY

There is abundant support from observations, simulations and theory that solar flares generate Alfvénic waves with enough energy density and Poynting flux to play a significant role in the flare. The likely sources of these waves are the fast magnetic reconnection process required in solar flares and braking of the sunward reconnection flow. Additionally, the sudden onset of the flare impulsive phase may launch Alfvénic fronts, that propagate downwards into the lower solar atmosphere. Once generated, Alfvénic waves evolve according to principles of field-aligned propagation, phase-mixing, instability and turbulent processes, with energy being lost to particle acceleration and/or heating the corona and lower atmosphere.

The last 15 years have brought the role of Alfvénic waves into considerably sharper focus. It is now widely acknowledged that they are likely to be a key component in the workings of solar flares, and nonthermal line widths have emerged as a key diagnostic that constrains wave energy density and Poynting flux. The above-the-loop region has been shown to be a key location for generating waves/turbulence, generation mechanisms have become addressable with high-performance 3D simulations, particle acceleration by Alfvénic waves/turbulence has gained greater observational support, and various investigations have shown the feasibility of waves directly heating in the lower atmosphere.

We finish with the message that key problems in flare physics are opening up, as modelling and theory embrace the inherent 3D and dynamic nature of real plasma and new observatories provide better windows on solar processes. These advances mean there is a realistic prospect of solving the grand challenges of fast reconnection, flare particle acceleration and flare energy transport within the span of a scientific career, and hence improving understanding and forecasting of space weather. Alfvénic waves are likely to be a central part of those solutions.

## ACKNOWLEDGEMENTS

AJBR thanks the many colleagues who have contributed to the development of the ideas presented in this article. The article benefited from discussions with participants of the 2023 AGU Chapman Conference on "Advances in Understanding Alfvén Waves in the Sun and the Heliosphere", and constructive comments and suggestions from two anonymous reviewers. I am grateful to creators and copyright holders for permitting the reproduction of figures in this work. Researching and writing the article made use of NASA's Astrophysics Data System Bibliographic Services.

DATA AVAILABILITY STATEMENT

No original data were processed in the creation of this review article.

Hertz, S. McKillop, S. Park, T. Perry, W. A. Podgorski, K. Reeves, S. Saar, P. Testa, H. Tian, M. Weber, C. Dunn, S. Eccles, S. A. Jaeggli, C. C. Kankelborg, K. Mashburn, N. Pust, L. Springer, R. Carvalho, L. Kleint, J. Marmie, E. Mazmanian, T. M. D. Pereira, S. Sawyer, J. Strong, S. P. Worden, M. Carlsson, V. H. Hansteen, J. Leenaarts, M. Wiesmann, J. Aloise, K. -C. Chu, R. I. Bush, P. H. Scherrer, P. Brekke, J. Martinez-Sykora, B. W. Lites, S. W. McIntosh, H. Uitenbroek, T. J. Okamoto, M. A. Gummin, G. Auker, P. Jerram, P. Pool, and N. Waltham. 2014. "The Interface Region Imaging Spectrograph (IRIS)." *Solar Physics* 289:2733-2779. doi: 10.1007/s11207-014-0485-y.

De Pontieu, Bart, Juan Martínez-Sykora, Paola Testa, Amy R. Winebarger, Adrian Daw, Viggo Hansteen, Mark C. M. Cheung, and Patrick Antolin. 2020. "The Multi-slit Approach to Coronal Spectroscopy with the Multi-slit Solar Explorer (MUSE)." *The Astrophysical Journal* 888:3. doi: 10.3847/1538-4357/ab5b03.

Dong, Chuanfei, Liang Wang, Yi-Min Huang, Luca Comisso, Timothy A. Sandstrom, and Amitava Bhattacharjee. 2022. "Reconnection-driven energy cascade in magnetohydrodynamic turbulence." *Science Advances* 8:eabn7627. doi: 10.1126/sciadv.abn7627.

Doschek, G. A. 1990. "Soft X-Ray Spectroscopy of Solar Flares: an Overview." *The Astrophysical Journal Supplement Series* 73:117. doi: 10.1086/191443.

Doschek, G. A., U. Feldman, K. P. Dere, G. D. Sandlin, M. E. Vanhoosier, G. E. Brueckner, J. D. Purcell, and R. Tousey. 1975. "Forbidden lines of highly ionized iron in solar flare spectra." *The Astrophysical Journal* 196:L83-L86. doi: 10.1086/181749.

Doschek, G. A., U. Feldman, R. W. Kreplin, and L. Cohen. 1980. "High-resolution X-ray spectra of solar flares. III - General spectral properties of X1-X5 type flares." *The Astrophysical Journal* 239:725-737. doi: 10.1086/158158.

Doschek, G. A., U. Feldman, J. F. Seely, and D. L. McKenzie. 1989. "High-Resolution X-Ray Spectra of Solar Flares. IX. Mass Upflow in the Long-Duration Flare of 1979 June 5." *The Astrophysical Journal* 345:1079. doi: 10.1086/167977.

Doschek, G. A., R. W. Kreplin, and U. Feldman. 1979. "High-resolution solar flare X-ray spectra." *The Astrophysical Journal* 233:L157-L160. doi: 10.1086/183096.

Doschek, G. A., D. E. McKenzie, and H. P. Warren. 2014. "Plasma Dynamics Above Solar Flare Soft X-Ray Loop Tops." *The Astrophysical Journal* 788:26. doi: 10.1088/0004-637x/788/1/26.

Doschek, G. A., J. F. Meekins, R. W. Kreplin, T. A. Chubb, and H. Friedman. 1971. "Iron-Line Emission during Solar Flares." *The Astrophysical Journal* 170:573. doi: 10.1086/151243.

Dudík, Jaroslav, Vanessa Polito, Elena Dzifčáková, Giulio Del Zanna, and Paola Testa. 2017. "Non-Maxwellian Analysis of the Transition-region Line Profiles Observed by the Interface Region Imaging Spectrograph." *The Astrophysical Journal* 842:19. doi: 10.3847/1538-4357/aa71a8.

Edwin, P. M., and B. Roberts. 1983. "Wave Propagation in a Magnetic Cylinder." *Solar Physics* 88:179-191. doi: 10.1007/bf00196186.

Emslie, A. G., and P. A. Sturrock. 1982. "Temperature minimum heating in solar flares by resistive dissipation of Alfvén waves." *Solar Physics* 80:99-112. doi: 10.1007/bf00153426.

Feldman, U., G. A. Doschek, R. W. Kreplin, and J. T. Mariska. 1980. "High-resolution X-ray spectra of solar flares. IV - General spectral properties of M type flares." *The Astrophysical Journal* 241:1175-1185. doi: 10.1086/158434.

Fermi, Enrico. 1949. "On the Origin of the Cosmic Radiation." *Physical Review* 75:1169-1174. doi: 10.1103/PhysRev.75.1169.

Fermo, R. L., J. F. Drake, and M. Swisdak. 2010. "A statistical model of magnetic islands in a current layer." *Physics of Plasmas* 17:010702. doi: 10.1063/1.3286437.

Fisher, G. H., D. J. Bercik, B. T. Welsch, and H. S. Hudson. 2012. "Global Forces in Eruptive Solar Flares: The Lorentz Force Acting on the Solar Atmosphere and the Solar Interior." *Solar Physics* 277:59-76. doi: 10.1007/s11207-011-9907-2.

Fisher, G. H., R. C. Canfield, and A. N. McClymont. 1985. "Flare loop radiative hydrodynamics. V - Response to thick-target heating. VI - Chromospheric evaporation due to heating by nonthermal